\documentclass[showpacs,preprint]{revtex4}

\usepackage{graphicx}
\usepackage{dcolumn}
\usepackage{amsmath}

\makeatletter
\def\btt#1{\texttt{\@backslashchar#1}}
\DeclareRobustCommand\bblash{\btt{\@backslashchar}} \makeatother

\begin{document}

\title[]{Rotating black hole and quintessence}
\author{Sushant~G.~Ghosh$^{a,\;b,}$} \email{sghosh2@jmi.ac.in, sgghosh@gmail.com}
\affiliation{$^{a}$ Astrophysics and Cosmology
Research Unit, School of Mathematics, Statistics and Computer Science, University of
KwaZulu-Natal, Private Bag 54001, Durban 4000, South Africa}
\affiliation{$^{b}$ Centre for Theoretical Physics, Jamia Millia
Islamia, New Delhi 110025, India}

\date{\today}
\begin{abstract}
We discuss spherically symmetric exact solutions of the Einstein equations for quintessential matter surrounding a black hole, which has an additional parameter ($\omega$) due to the quintessential matter, apart from the mass ($M$).  In turn, we employ the Newman\(-\)Janis complex transformation to this spherical quintessence black hole solution and present a rotating counterpart that is identified, for $\alpha=-e^2 \neq 0$ and $\omega=1/3$, exactly as the Kerr\(-\)Newman black hole, and as the Kerr black hole when $\alpha=0$. Interestingly, for a given value of parameter $\omega$, there exists a critical rotation parameter ($a=a_{E}$), which corresponds to an extremal black hole with degenerate horizons, while for $a<a_{E}$, it describes a non-extremal black hole with Cauchy and event horizons, and no black hole for $a>a_{E}$. We find that the extremal value $a_E$ is also influenced by the parameter $\omega$ and so is the \textit{ergoregion}.
\end{abstract}{\large }



\maketitle
The accelerating expansion of the Universe implies the crucial contribution of matter with negative pressure
to the evolution of the Universe, which could be the cosmological constant or so-called quintessence matter. If quintessence matters exists all over in the Universe, it can also be around a black hole. In this letter, we are interested in getting the rotating counterpart of the solution to the Einstein equations obtained with the assumption of spherical symmetry, with the quintessence  matter as a source of energy\(-\)momentum obtained by Kislev \cite{Kiselev} and was also  rigorously analyzed by himself and others \cite{Kiselev,Chen:2005qh,Zhang:2006ij,Chen:2008ra,Wei:2011za,Thomas:2012zzc,FERNANDO:2013uxa,Tharanath:2013jt}. Let us commence with the general spherically symmetric spacetime
\begin{equation}\label{metric}
ds^2=g_{a b} \otimes dx^{a} \otimes dx^{b},\;\;\;\;(a,b=0,1,2,3),
\end{equation}
with $g_{a b}= \text{diag} (-f(r),f(r)^{-1},r^2,r^2\sin^2 \theta)$,
the energy\(-\)momentum tensor for the quintessence \cite{Kiselev} is given by
\begin{eqnarray}\label{emt}
T^t_t=T^r_r = \rho_q, \nonumber \\
T^{\theta}_{\theta}=T^{\phi}_{\phi} = - \frac{1}{2} \rho_q (3 \omega+1).
\end{eqnarray}
On using the Einstein equations $G_{ab} = T_{ab}$, one obtains
\begin{equation}\label{sol}
f(r) = 1- \frac{r_g}{r} + \frac{\beta}{r^{3\omega+1}},
\end{equation}
where $\beta$ and $r_g$ are the normalization factor. The density of quintessence matter $\rho_q$ is given by
\begin{equation}\label{density}
\rho_q =  \frac{\beta}{2} \frac{3 \omega}{r^{3 (\omega +1)}}.
\end{equation}
The sign of the normalization constant should coincide with the sign of the matter state parameter, i.e.
$\beta \omega \geq 0$  implying that $\beta $  is negative for the quintessence and hence we choose $\alpha = -\beta$.
Thus a metric of exact spherically symmetric solutions for the Einstein equations describing black holes surrounded by quintessential matter with the energy\(-\)momentum tensor (\ref{emt}) is given by
\begin{equation}ds^2 =  \left[ 1-  \frac{2 M}{r} - \frac{\alpha}{r^{3\omega+1}}\right]
dt^2 - \frac{dr^2}{\left[ 1-  \frac{2 M}{r} - \frac{\alpha}{r^{3\omega+1}}\right]} - {r^2d\Omega^2}.
\label{solA}
\end{equation}
Here $r_g$ is related to the   black hole mass via $r_g=2M$, and $\omega$ is the quintessential state parameter.
The Ricci scalars for the solution reads
\begin{equation}
R=R^{ab}_{ab} = \frac{9 \alpha^2 \omega^2 (9 \omega^2 + 6 \omega +3)}{2 r^{2(3\omega+1)}},
\end{equation}
indicating scalar polynomial singularity at $r=0$ if $\omega \neq \{0, \frac{1}{3},-1\}$. Thus we have  a general form of exact spherically symmetric solutions for the Einstein equations describing black holes surrounded by quintessential matter. The parameter $ \omega $ has to have the range, $-1 < \omega < -1/3$ for a de Sitter horizon and which causes the acceleration, and $-1/3 < \omega <0$ for the asymptotically flat solution. It is the most general spherically symmetric static solution of Einstein's field equation coupled with quintessence matter as a source. For $ \alpha = 0$, it reduces to the Schwarzschild solution. The case with the relativistic matter state parameter $\omega =1/3$, with $ \alpha = -e^2$, corresponds to Reissner\(-\)Nordstrom black hole with
\begin{equation}\label{sol}
f(r) = 1- \frac{r_g}{r} + \frac{e^2}{r^2}.
\end{equation}
The solution for the Reissner\(-\)Nordstrom black hole surrounded by the
quintessence gives
\begin{equation}\label{solRN}
f(r) = 1- \frac{r_g}{r} + \frac{e^2}{r^2}- \frac{\alpha}{r^{3\omega+1}}.
\end{equation}
The borderline case
of $ \omega = - 1 $ of the extraordinary quintessence covers the cosmological constant term, and spacetime (\ref{solA}) reduces to the Schwarzschild de Sitter black hole.

The purpose of this letter is to seek the generalization of the solution (\ref{solA}) to the axially symmetric case or to the Kerr\(-\)like metric. The  Kerr metric \cite{kerr} is beyond question the most extraordinary exact solution in the Einstein theory of general relativity, which represents the prototypic black hole that can arise from gravitational collapse, which contains an event horizon \cite{bc}. It is well known that Kerr black hole enjoy many interesting
properties distinct from its non-spinning counterpart, i.e., from Schwarzschild black hole. However, there is a surprising
connection between the two black holes of Einstein theory, and was analyzed by Newman and Janis \cite{nja,ncept,etn,etn1} that the Kerr metric \cite{kerr} could be obtained from the Schwarzschild metric using a complex transformation within the framework of the
Newman\(-\)Penrose formalism \cite{np}. A similar procedure was applied to the Reissner\(-\)Nordström metric to generate the previously unknown Kerr\(-\)Newman metric \cite{ncept}. It is an ingenious algorithm to construct a metric for rotating black hole from static spherically symmetric solutions in  Einstein gravity. The Newman\(-\)Janis method has proved to be prosperous in generating new stationary solutions of the Einstein field equations and the  method have also been studied outside the domain of general relativity \cite{Capo,Qw,Bambi,Ghosh:2014hea,Toshmatov:2014nya,Larranaga:2014uca,Ghosh:2013bra,Neves:2014aba,Ghosh:2014pba,Azreg-Ainou:2014nra}, although it may lead to additional stresses \cite{Neves:2014aba,Carmeli,Ghosh:2014hea,Qw}. For possible physical interpretations of the algorithm, see \cite{ejf,ejf1}, and for discussions on more general complex transformations, see \cite{ejf,ejf1,ds}. For a review on the Newman\(-\)Janis algorithm see, e.g., \cite{Rd}.

Next, we wish to derive a rotating analogue of the static spherically symmetric quintessence solution (\ref{solA}) by employing the Newman\(-\)Janis \cite{nja} complex transformation. To attempt the similar for static quintessence solution (\ref{solA}) to generate rotating counterpart, we take the quintessence solution  (\ref{solA}) and perform the Eddington\(-\)Finkelstein coordinate
transformation, $$du = dt - \left[ 1-  \frac{2 M}{r} - \frac{\alpha}{r^{3\omega+1}}\right] dr,$$ so that the metric takes the form
\begin{eqnarray}
ds^2 =  \left[ 1-  \frac{2 M}{r} - \frac{\alpha}{r^{3\omega+1}}\right] du^2 - 2 dudr - {r^2}d\Omega^2. \label{SchwEF1}
\end{eqnarray}
Following the Newman\(-\)Janis prescription \cite{nja, Capo}, we can write the metric in terms of null tetrad, $ Z^a = (l^a,\;n^a,\;m^a,\;\bar{m}^a$), as
\begin{equation}
{g}^{ab} = l^a n^b +  l^b n^a - m^a \bar{m}^b-\bar{m}^a {m}^b,
\label{NPmetric}
\end{equation}
where null tetrad are
\begin{eqnarray*}
l^a &=& \delta^a_r,\\
n^a &=&  \delta^a_u - \frac{1}{2} \left[ 1-  \frac{2 M}{r} - \frac{\alpha}{r^{3\omega+1}}\right]\delta^a_r ,\\
 m^a &=& \frac{1}{\sqrt{2}r}   \left( \delta^a_{\theta}
  + \frac{i}{\sin\theta} \delta^a_{\phi} \right).
\end{eqnarray*}
The null tetrad are orthonormal and obey the metric conditions
\begin{eqnarray}
l_{a}l^{a} = n_{a}n^{a} = ({m})_{a} ({m})^{a} = (\bar{m})_{a} (\bar{m})^{a}= 0,  \nonumber \\
l_{a}({m})^{a} = l_{a}(\bar{m})^{a} = n_{a}({m})^{a} = n_{a}(\bar{m})^{a}= 0, \; \nonumber \\
l_a n^a = 1, \; ({m})_{a} (\bar{m})^{a} = 1,
\end{eqnarray}
 Now we allow for some $r$ factor in the null vectors to take on complex values. We rewrite the null vectors in the form \cite{Capo,Qw}
\begin{eqnarray*}\label{NPjnw}
l^a &=& \delta^a_r, \\
n^a &=& \left[ \delta^a_u - \frac{1}{2} \left[ 1-  {M} \left(\frac{1}{r}+\frac{1}{\bar{r}}\right) -\frac{\alpha}{{(r\bar{r})}^{\frac{3\omega+1}{2}}} \right] \delta^a_r \right], \\
m^a &=& \frac{1}{\sqrt{2} \bar{r}}  \left( \delta^a_{\theta}
  + \frac{i}{\sin\theta} \delta^a_{\phi} \right).
\end{eqnarray*}
with $\bar{r}$ being the complex conjugate of $r$.  Following the Newman\(-\)Janis prescription
\cite{nja}, we now write,
\begin{equation}\label{transf}
{x'}^{a} = x^{a} + ia (\delta_r^{a} - \delta_u^{a})
\cos\theta \rightarrow \\ \left\{\begin{array}{ll}
u' = u - ia\cos\theta, \\
r' = r + ia\cos\theta, \\
\theta' = \theta, \\
\phi' = \phi. \end{array}\right.
\end{equation}
where $a$ is the rotation parameter. Simultaneously, let null tetrad vectors $Z^a$ undergo a
transformation $Z^a = Z'^a{\partial x'^a}/{\partial x^b} $ in the
usual way, we obtain
\begin{eqnarray}\label{NPkerr}
l^a &=& \delta^a_r, \nonumber \\
n^a &=& \left[ \delta^a_u - \frac{1}{2} \left[1 - \frac{2Mr}{\Sigma } - \frac{\alpha}{{\Sigma}^{\frac{3\omega+1}{2}}} \right] \delta^a_r \right], \nonumber \\
 m^a &=& \frac{1}{\sqrt{2}(r+ia\cos\theta)}  \nonumber \\
 & & \times  \left(ia(\delta^a_u-\delta^a_r)\sin\theta + \delta^a_{\theta} + \frac{i}{\sin\theta} \delta^a_{\phi} \right),
 \end{eqnarray}
where $\rho=r^2+a^2\cos \theta$ and we have dropped the primes.  Using tetrad (\ref{NPkerr}), the non zero component of the inverse of a new metric can be written as
\begin{eqnarray}\label{KCinvm}
&& g^{u u } = -\frac{a^2 \sin^2(\theta )}{ \Sigma (r ,\theta )} \,\,\, , \,\,
g^{u \phi} = -\frac{a}{  \Sigma  (r ,\theta )} , \nonumber  \\
&& g^{\phi \phi} = -\frac{1}{   \Sigma  (r, \theta ) \sin^2 \theta } \,\,\, , \,\,
g^{\theta \theta} = -\frac{1}{  \Sigma  (r,\theta )},  \nonumber \\
&& g^{rr} = -\frac{a^2 \sin ^2 \theta }{  \Sigma  (r, \theta)} - \zeta(r,\theta) \,\,\, , \,\,
g^{r \phi} = \frac{a}{ \Sigma  (r, \theta)},
\nonumber  \\
&& g^{u r } = \frac{a^2 \sin ^2(\theta )}{  \Sigma  (r, \theta)} + 1,
\end{eqnarray}
where,
\begin{equation}
\zeta(r,\theta) = 1 - \frac{2Mr}{\Sigma } - \frac{\alpha}{{\Sigma}^{\frac{3\omega+1}{2}}}.
\end{equation}
\begin{figure*}
	\begin{tabular}{c c}
		\includegraphics[scale=0.62]{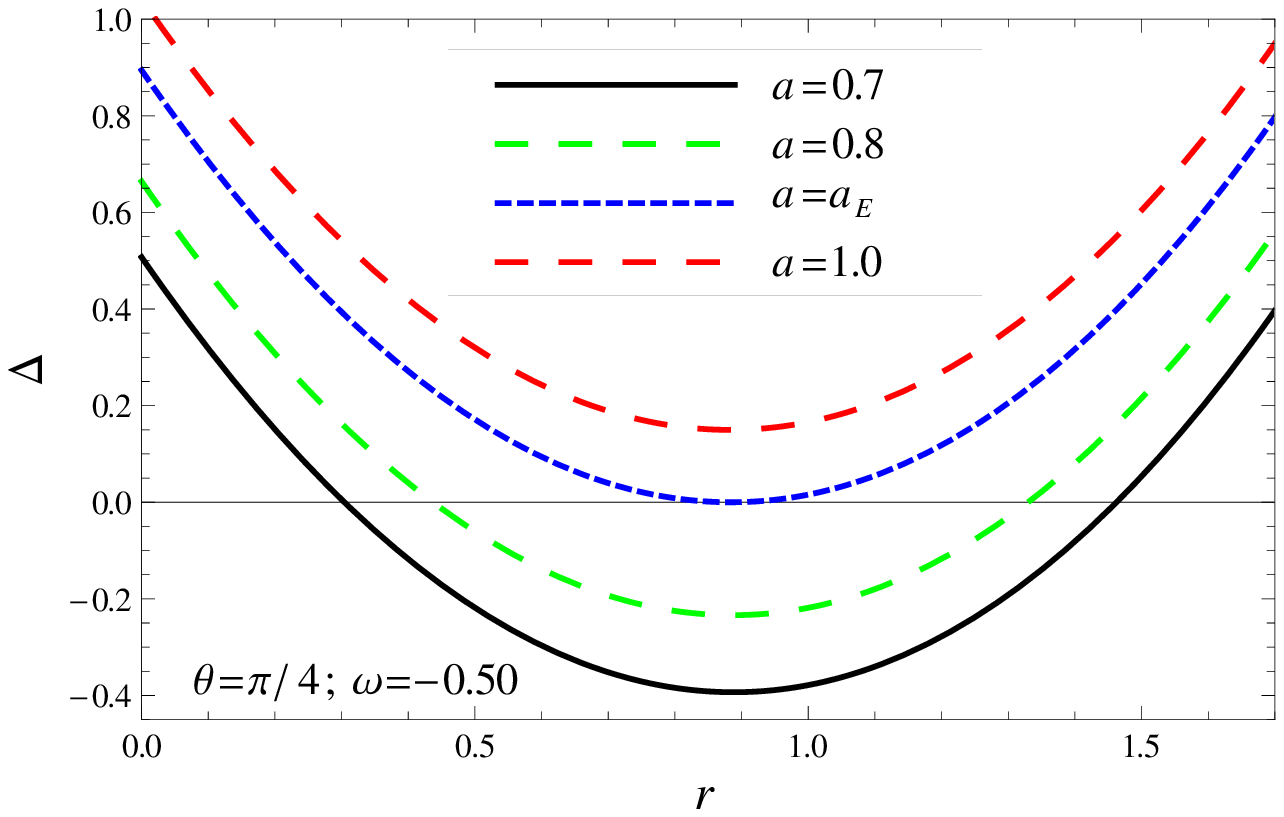}\hspace{-0.7cm}
	   &\includegraphics[scale=0.62]{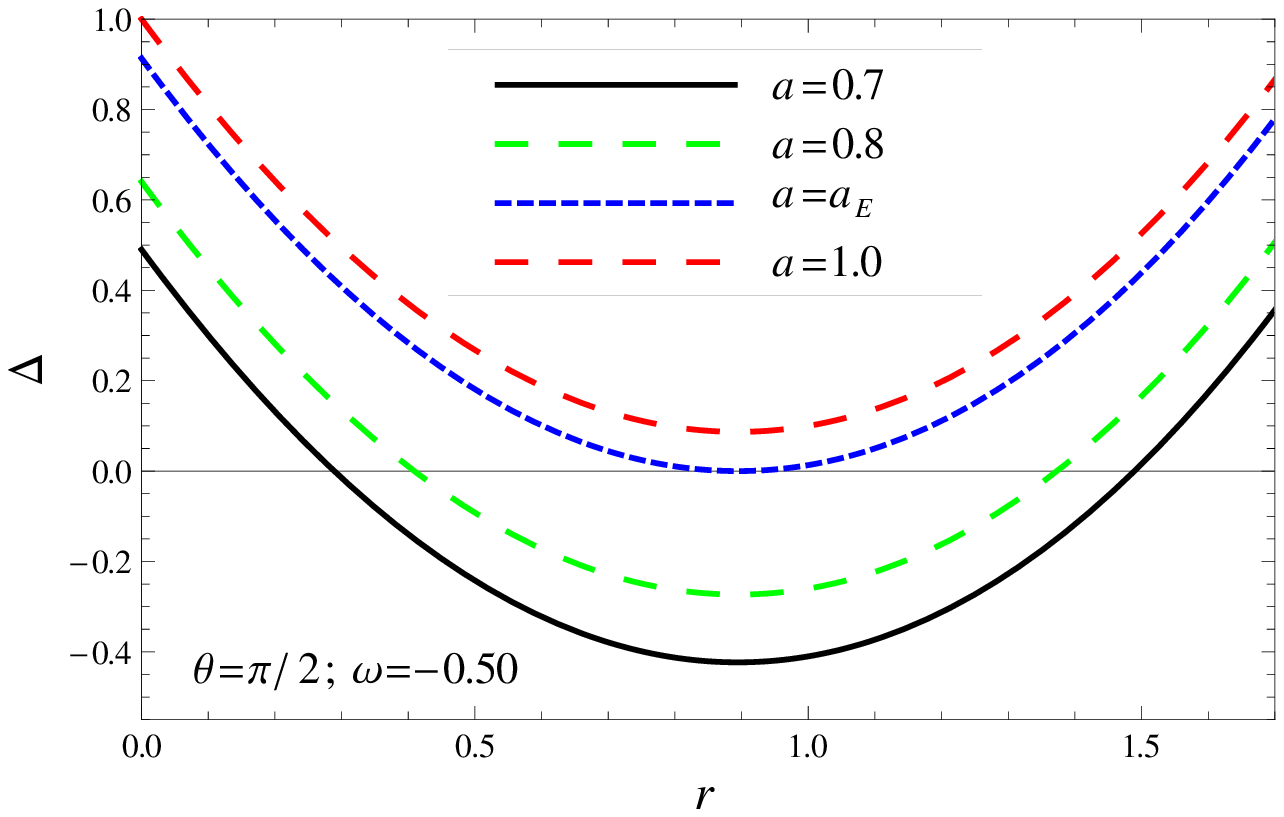}\\
	    \includegraphics[scale=0.62]{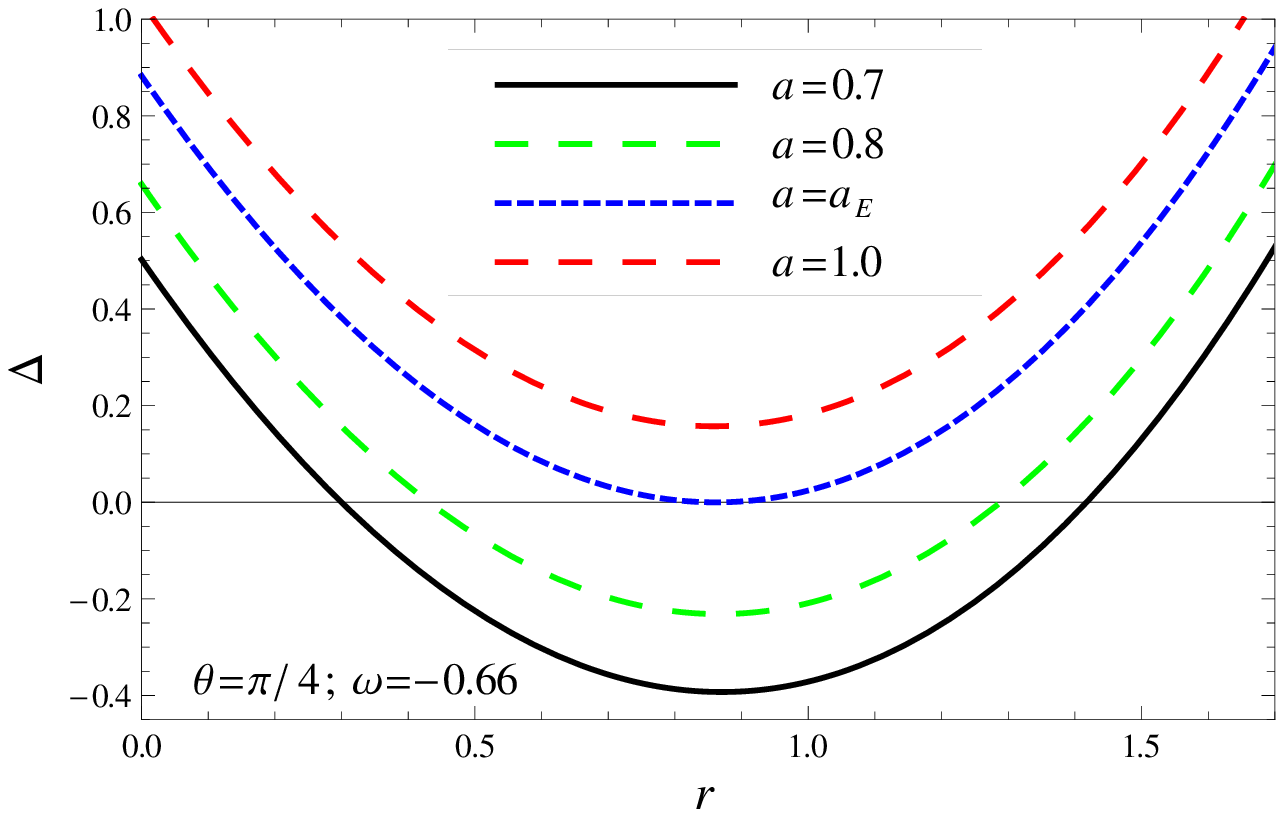}\hspace{-0.7cm}
	   &\includegraphics[scale=0.62]{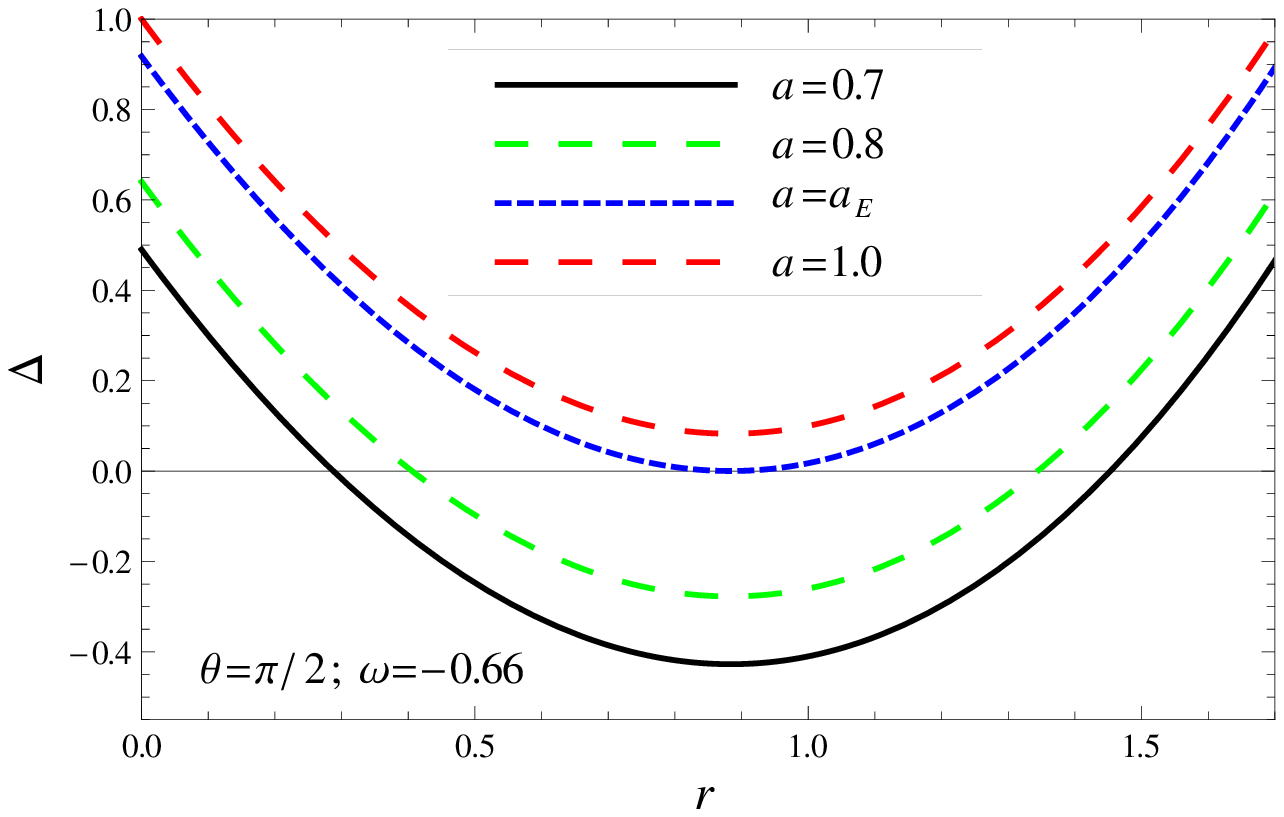}\\
	    \includegraphics[scale=0.62]{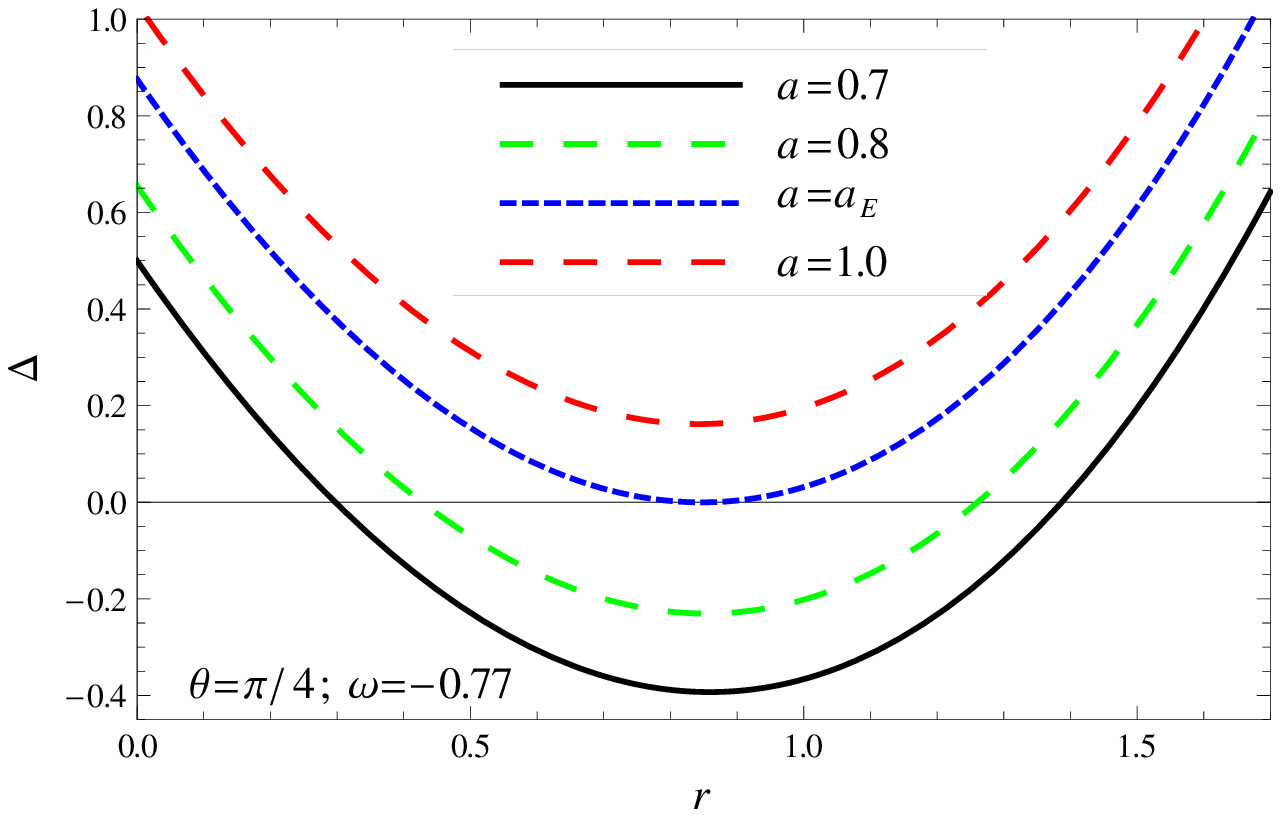}\hspace{-0.7cm}
	   &\includegraphics[scale=0.62]{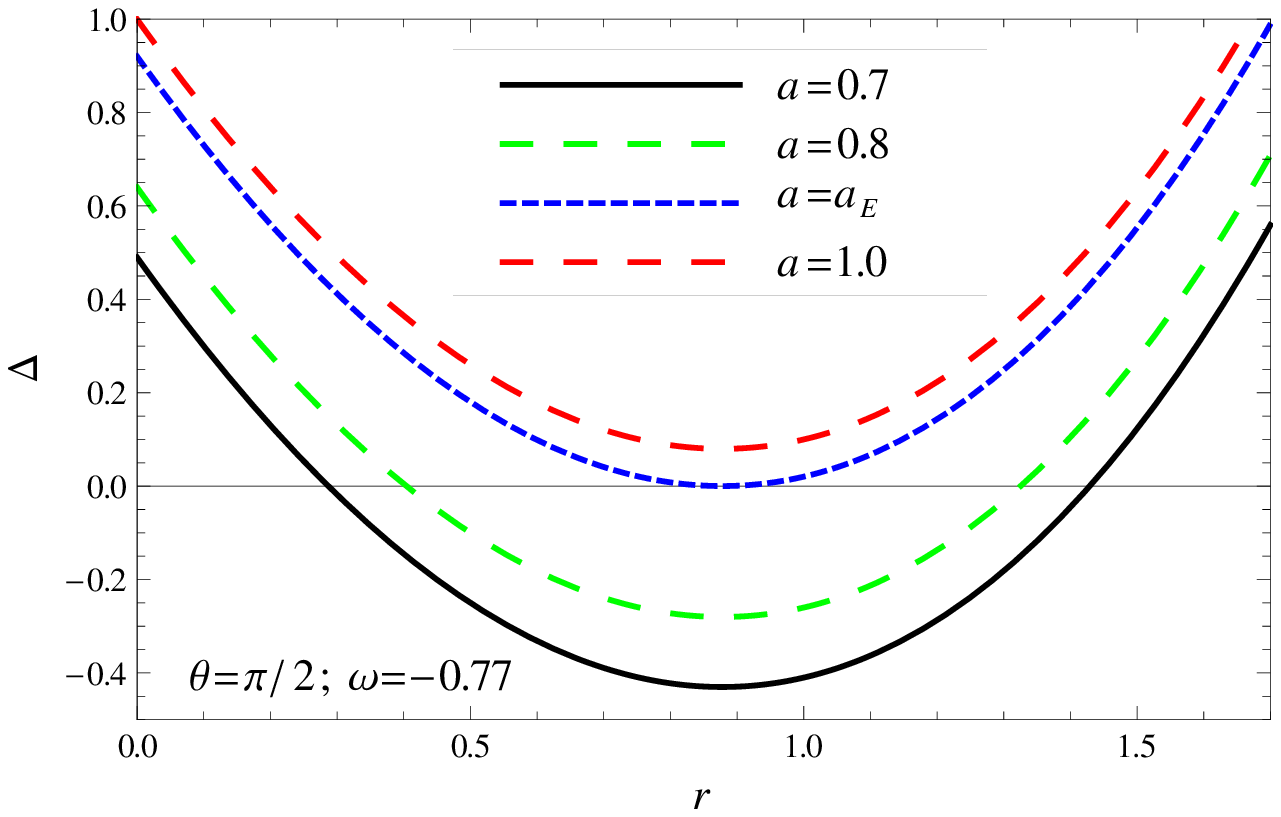}\\
	    \includegraphics[scale=0.62]{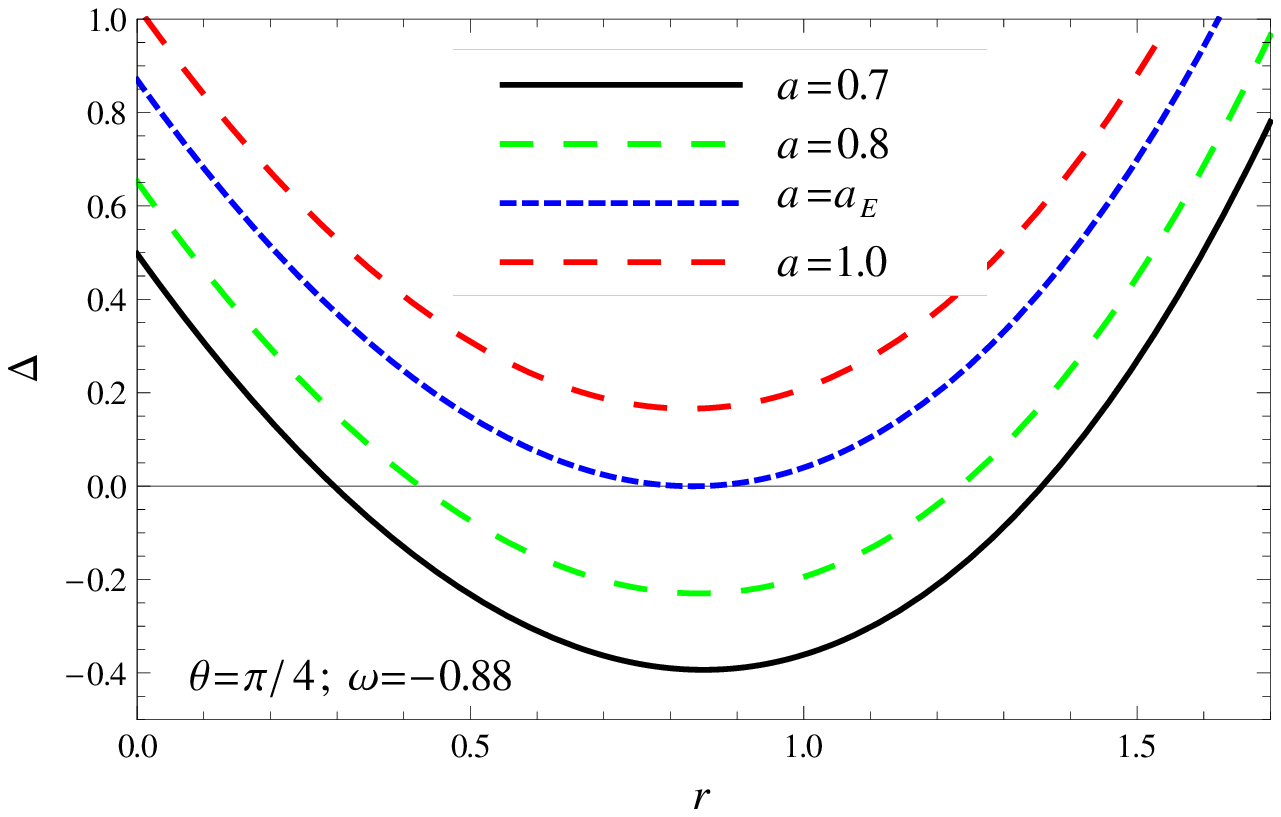}\hspace{-0.7cm}
	   &\includegraphics[scale=0.62]{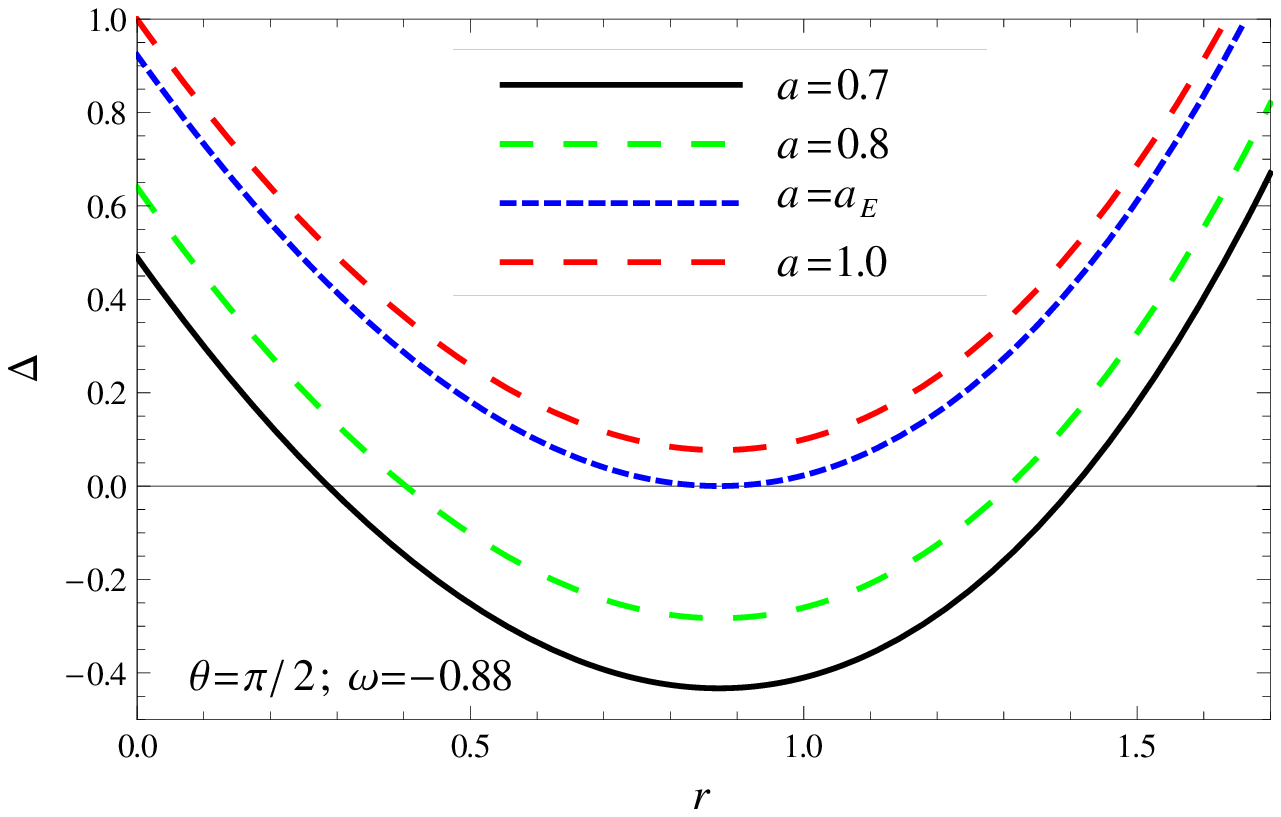}
		\end{tabular}
	\caption{Plot showing the behavior of $\Delta$ vs. $r$ for fixed values of $\alpha=-0.1$, and $M=1$ by varying $a$. The case $a=a_E$ corresponds to an extremal black hole}\label{fig1}
\end{figure*}

\begin{figure*}
	\begin{tabular}{c c}
		\includegraphics[scale=0.62]{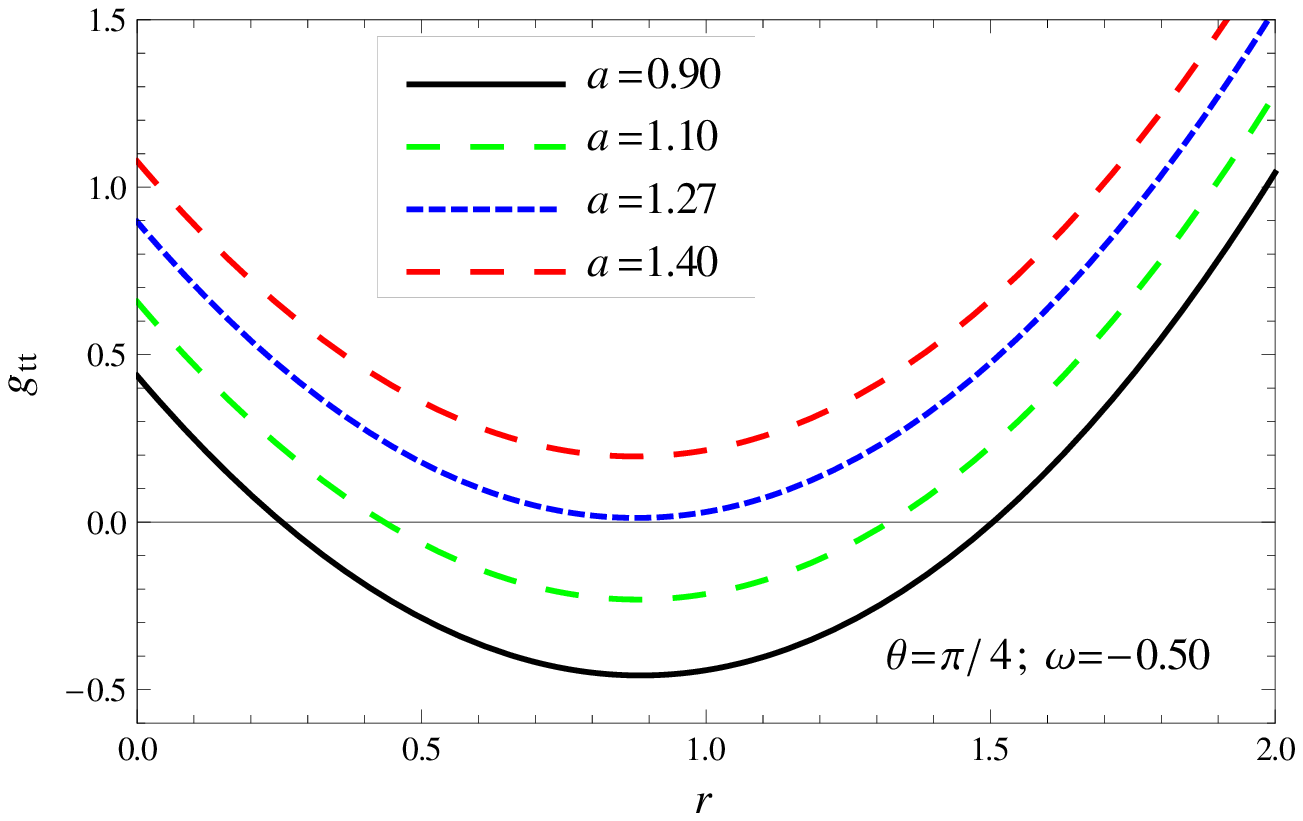}\hspace{-0.7cm}
	   &\includegraphics[scale=0.62]{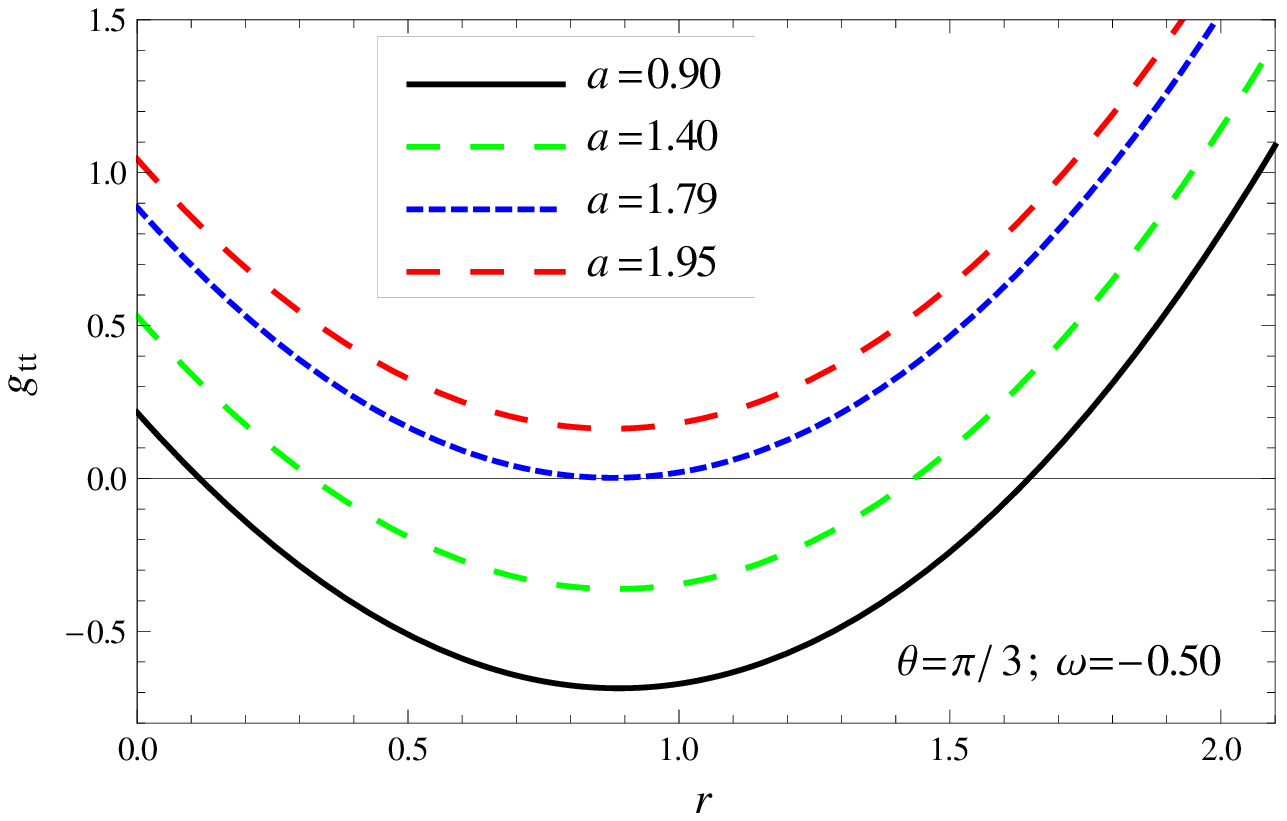}\\
		\includegraphics[scale=0.62]{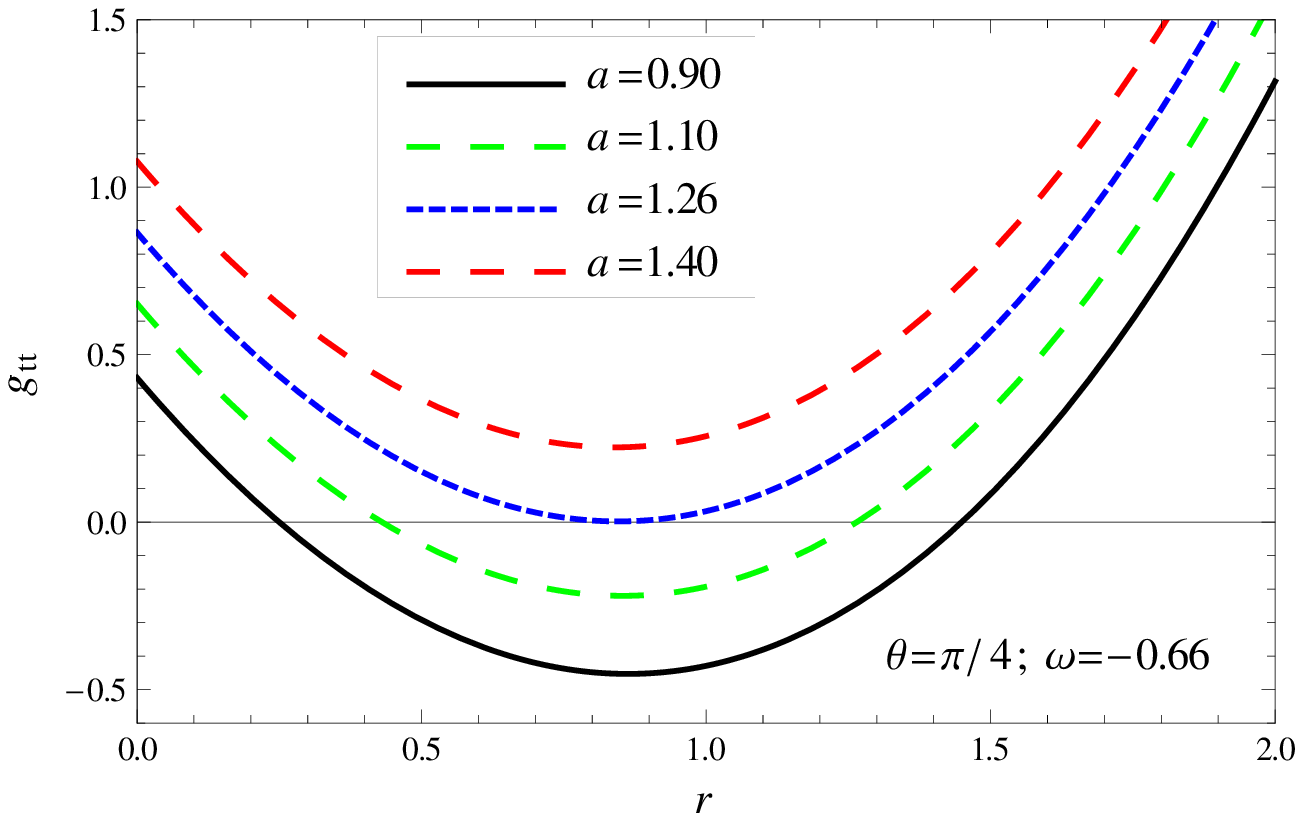}\hspace{-0.7cm}
	   &\includegraphics[scale=0.62]{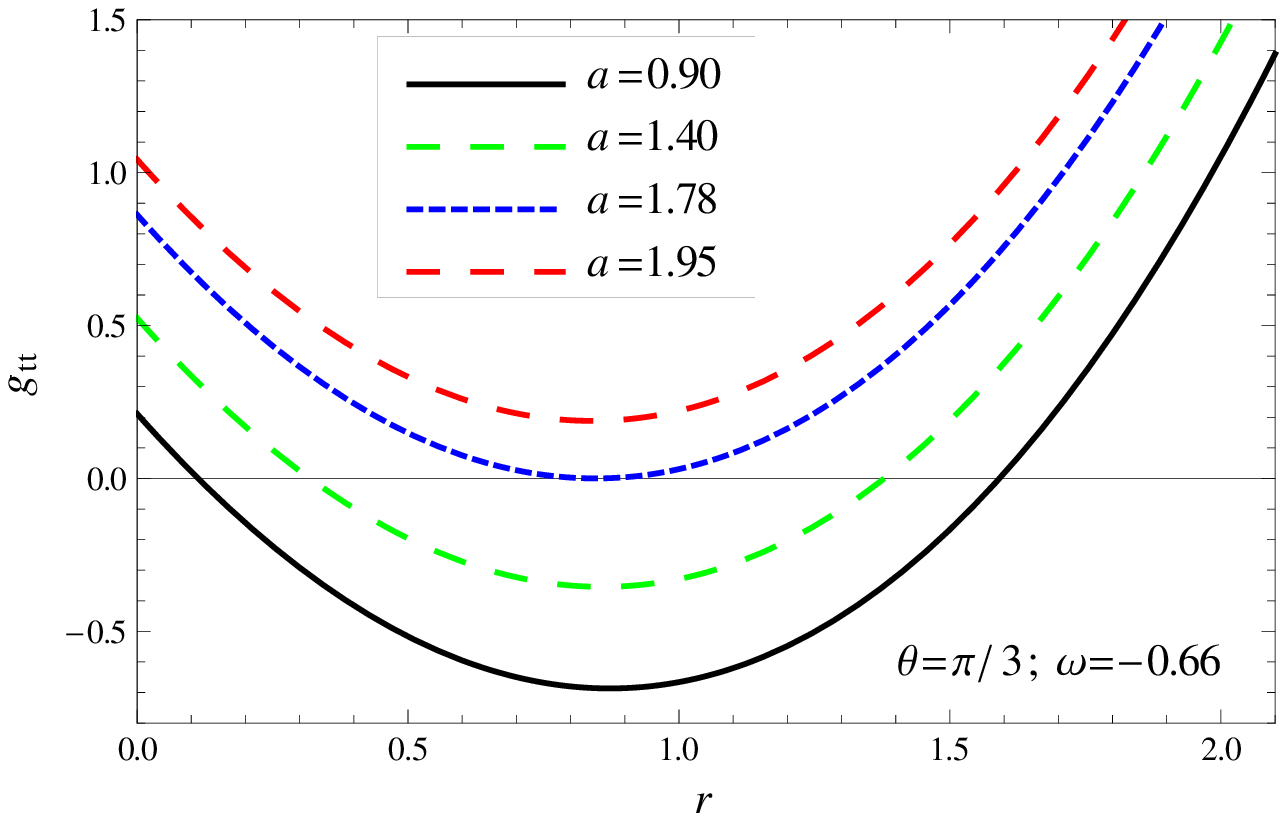}\\
	    \includegraphics[scale=0.62]{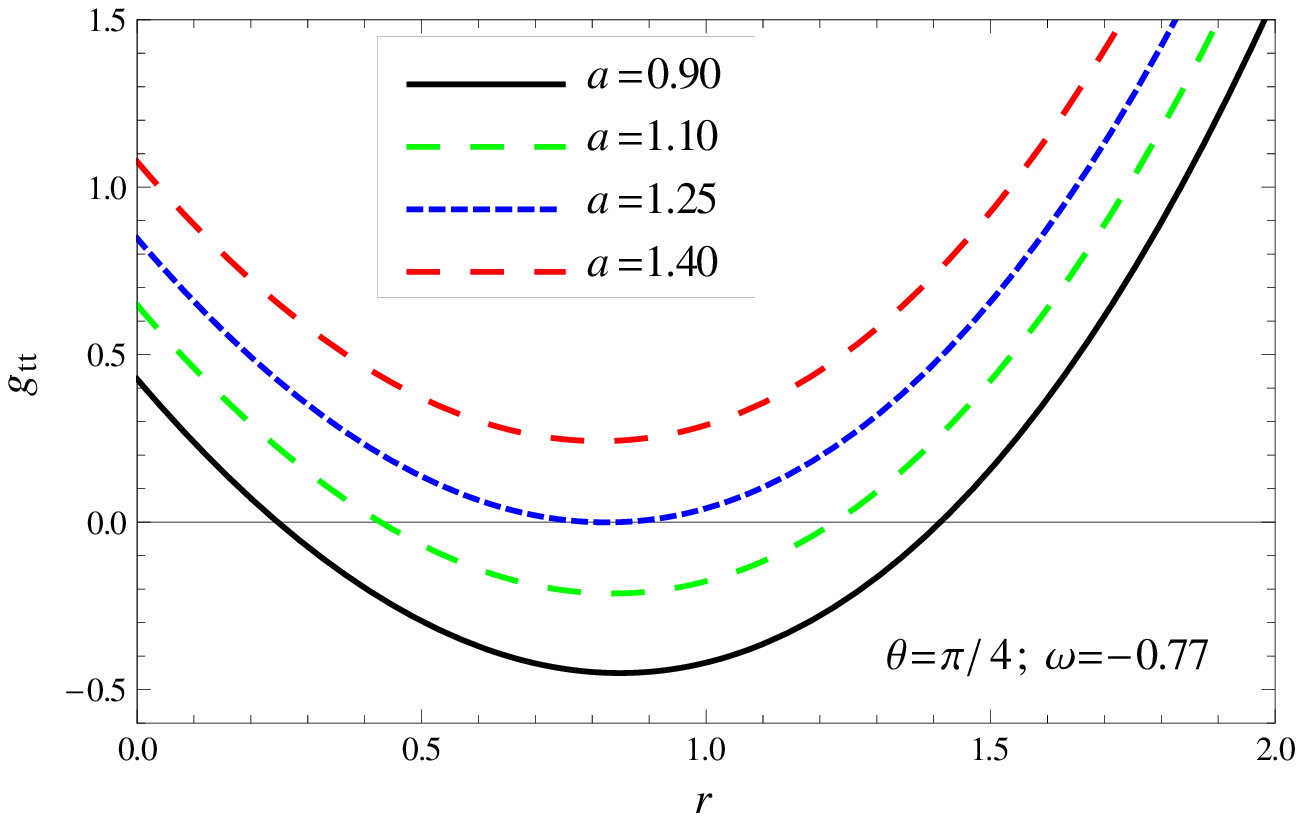}\hspace{-0.7cm}
	   &\includegraphics[scale=0.62]{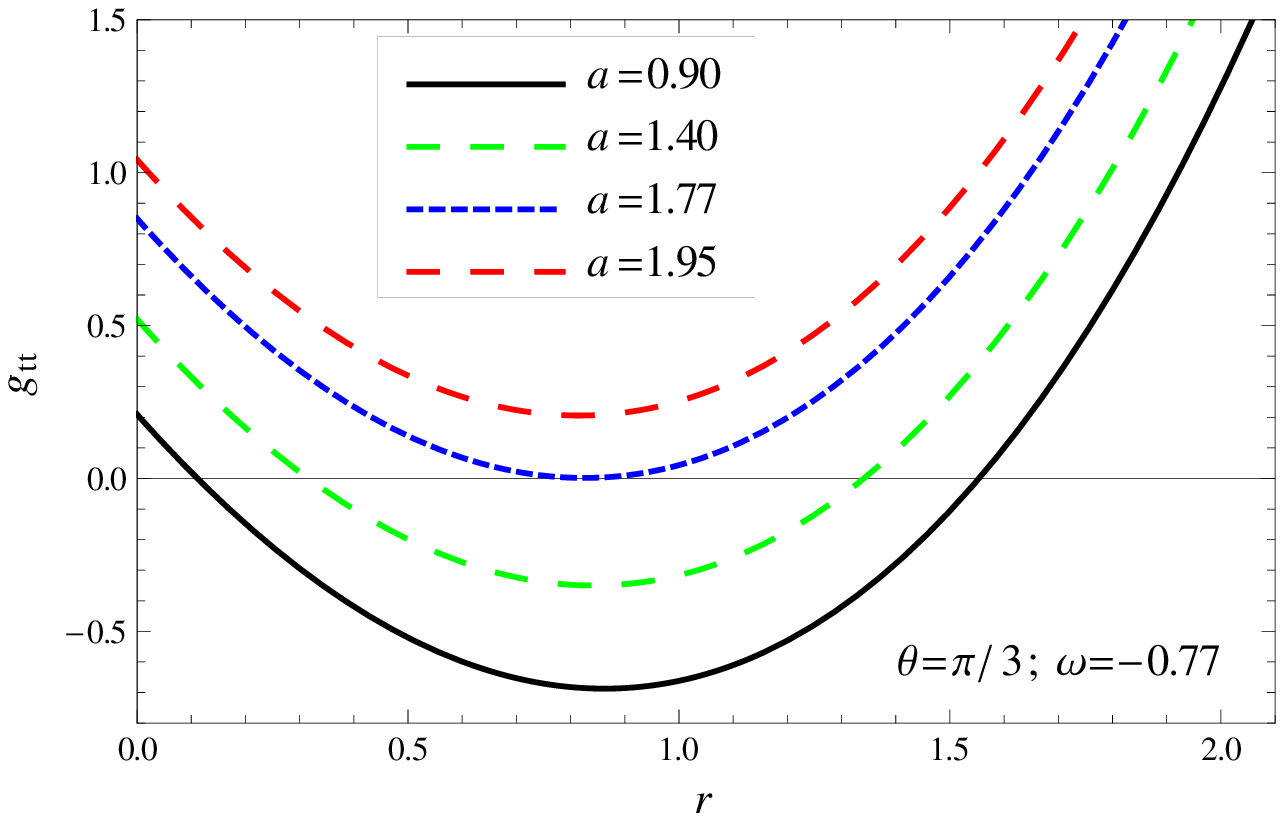}\\
	    \includegraphics[scale=0.62]{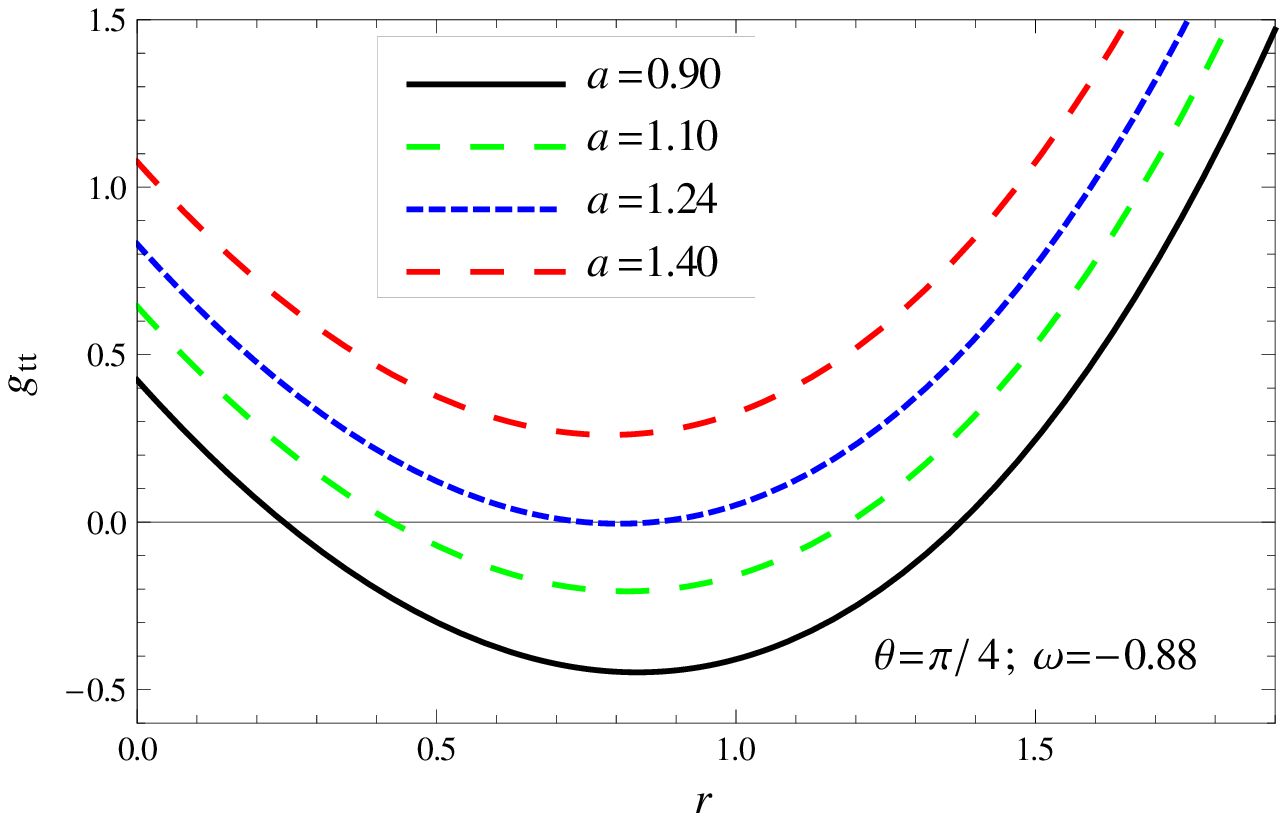}\hspace{-0.7cm}
	   &\includegraphics[scale=0.62]{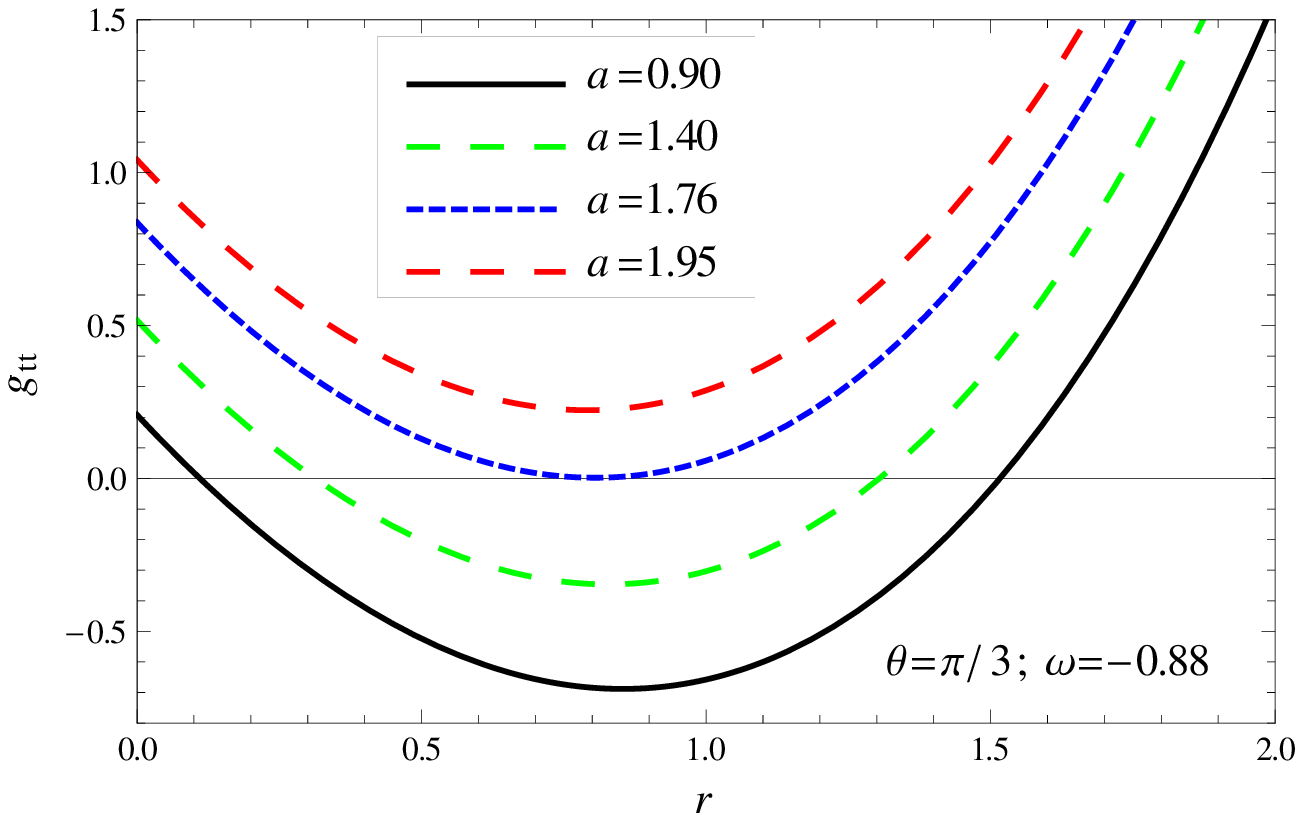}
	\end{tabular}
	\caption{Plot showing the behavior of $g_{tt}$ vs. $r$ for fixed values of $\alpha=-1$ and $M=1$ by varying $a$}\label{fig2}
\end{figure*}

\begin{figure*}
	\begin{tabular}{c c c c}
		\includegraphics[scale=0.42]{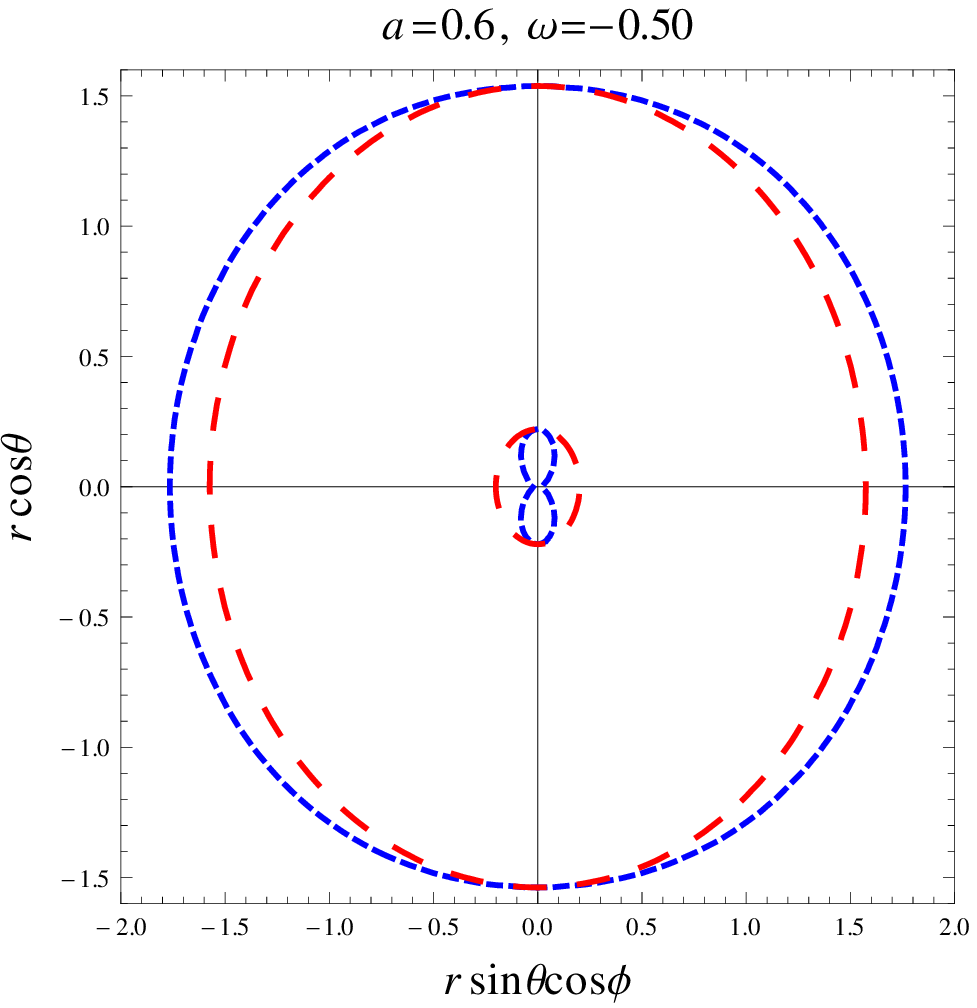}\hspace{-0.2cm}
		\includegraphics[scale=0.42]{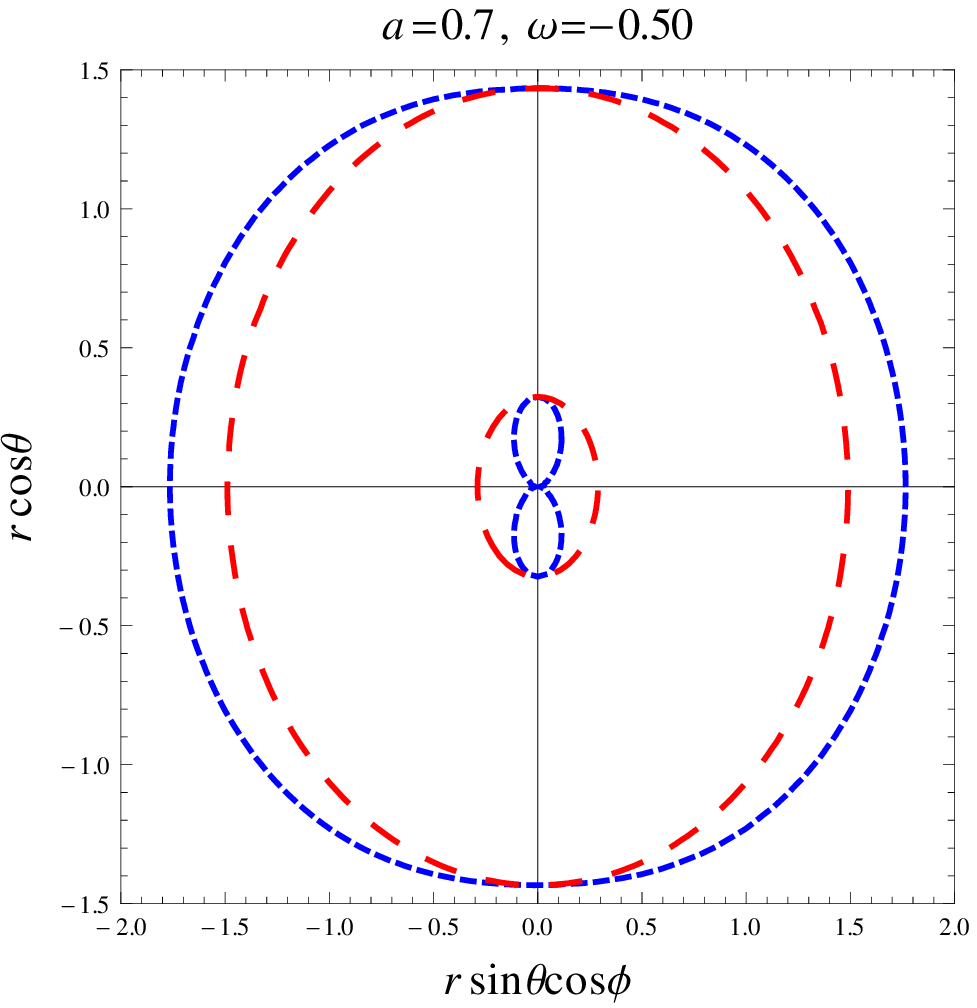}\hspace{-0.2cm}
		\includegraphics[scale=0.42]{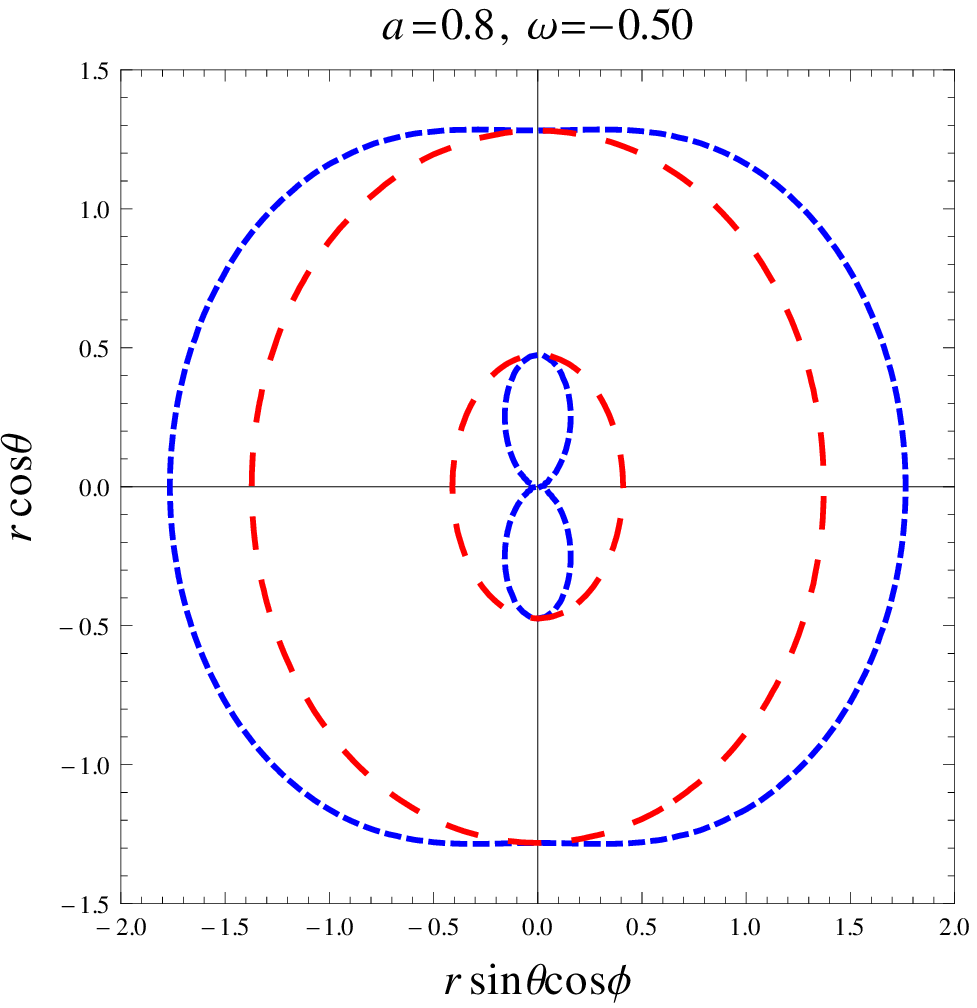}\hspace{-0.2cm}
	   &\includegraphics[scale=0.42]{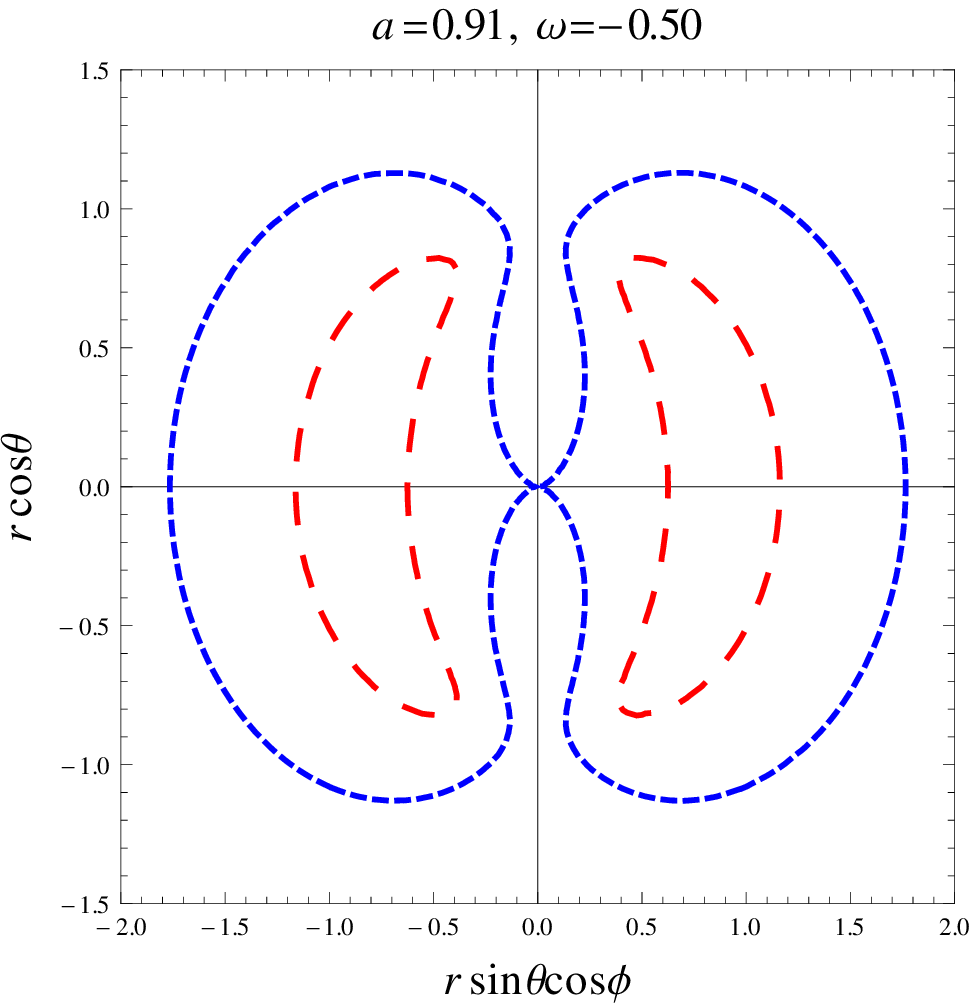}\\
		\includegraphics[scale=0.42]{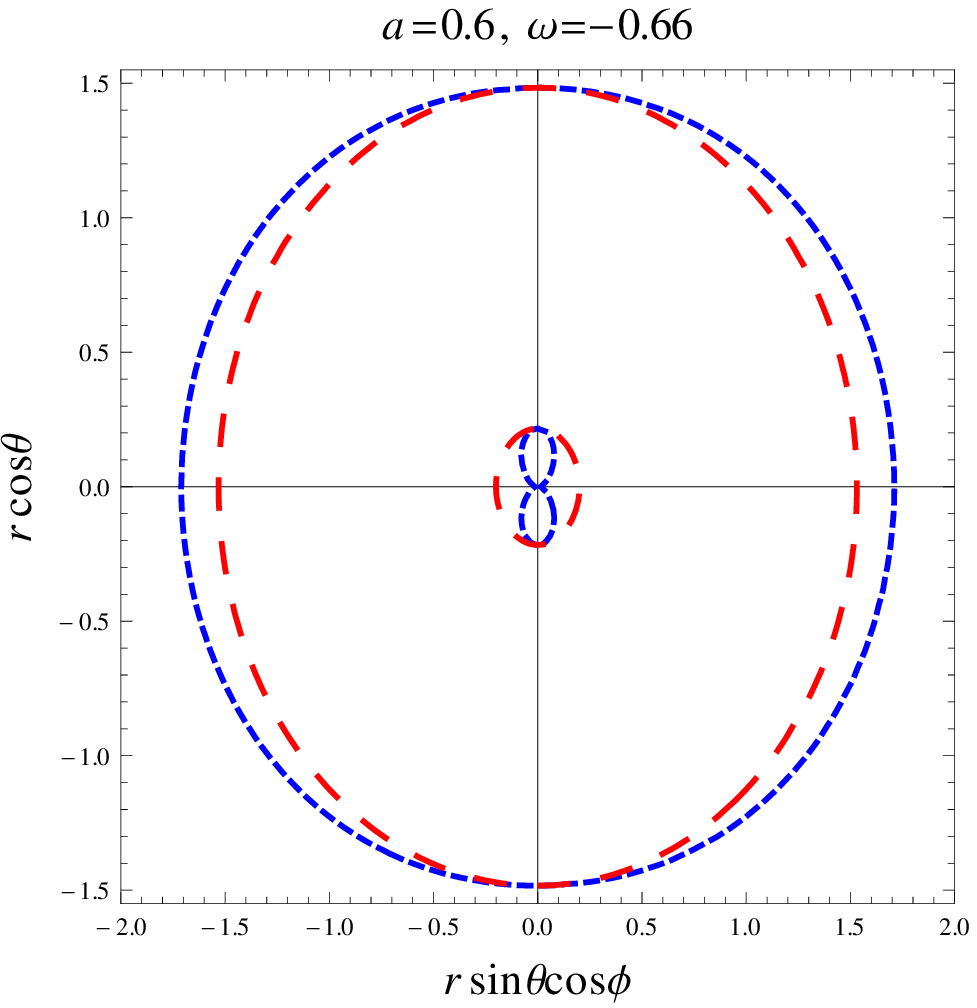}\hspace{-0.2cm}
		\includegraphics[scale=0.42]{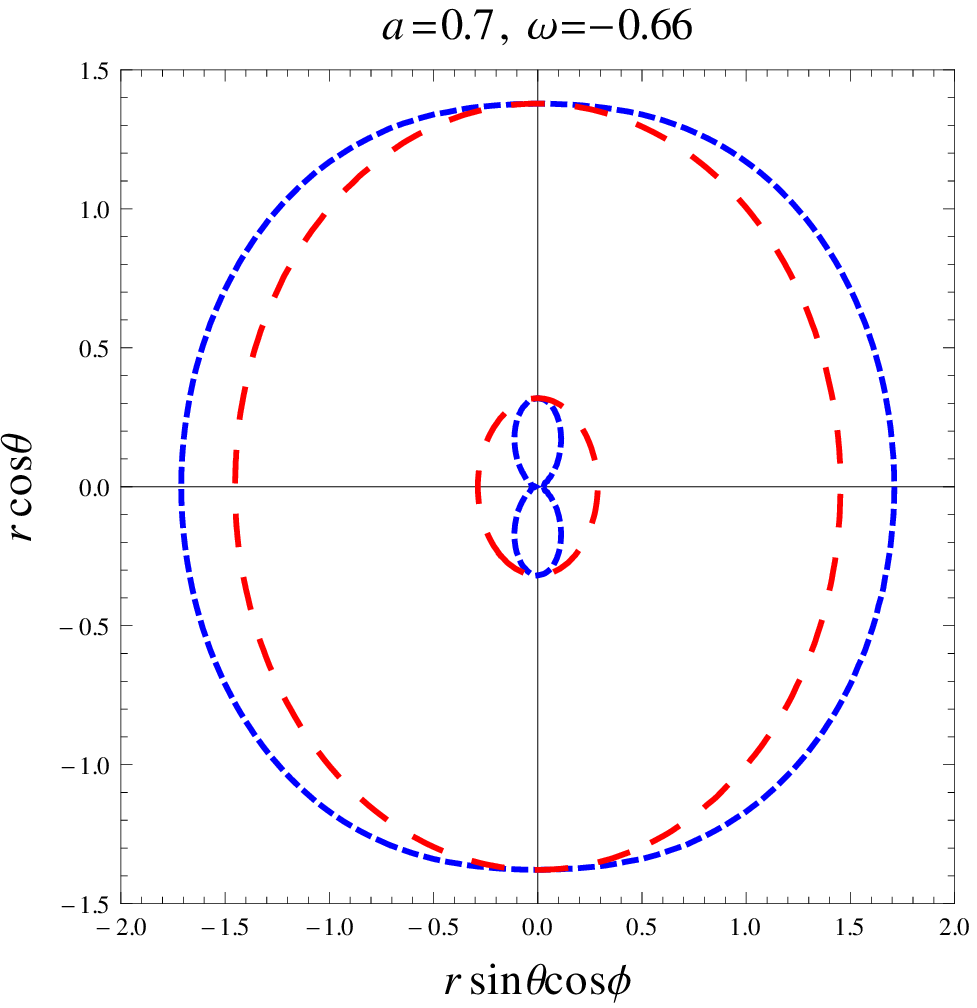}\hspace{-0.2cm}
		\includegraphics[scale=0.42]{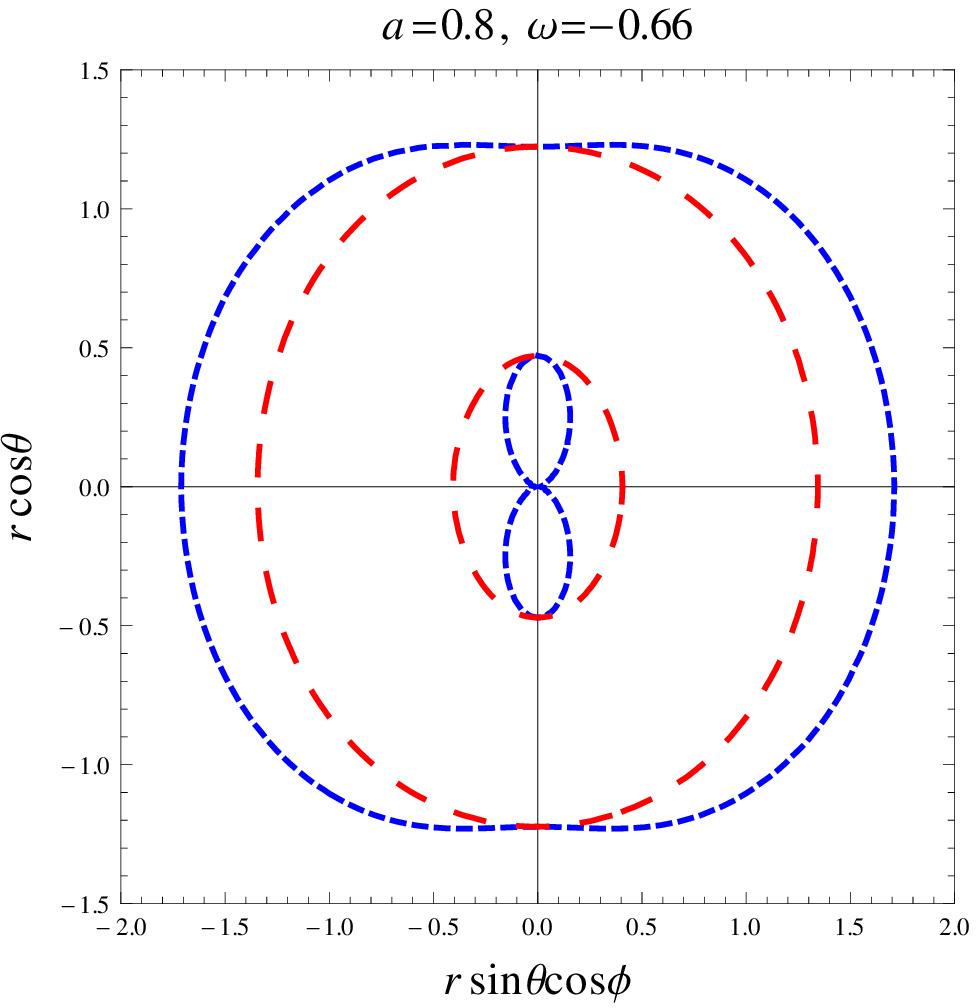}\hspace{-0.2cm}
	   &\includegraphics[scale=0.42]{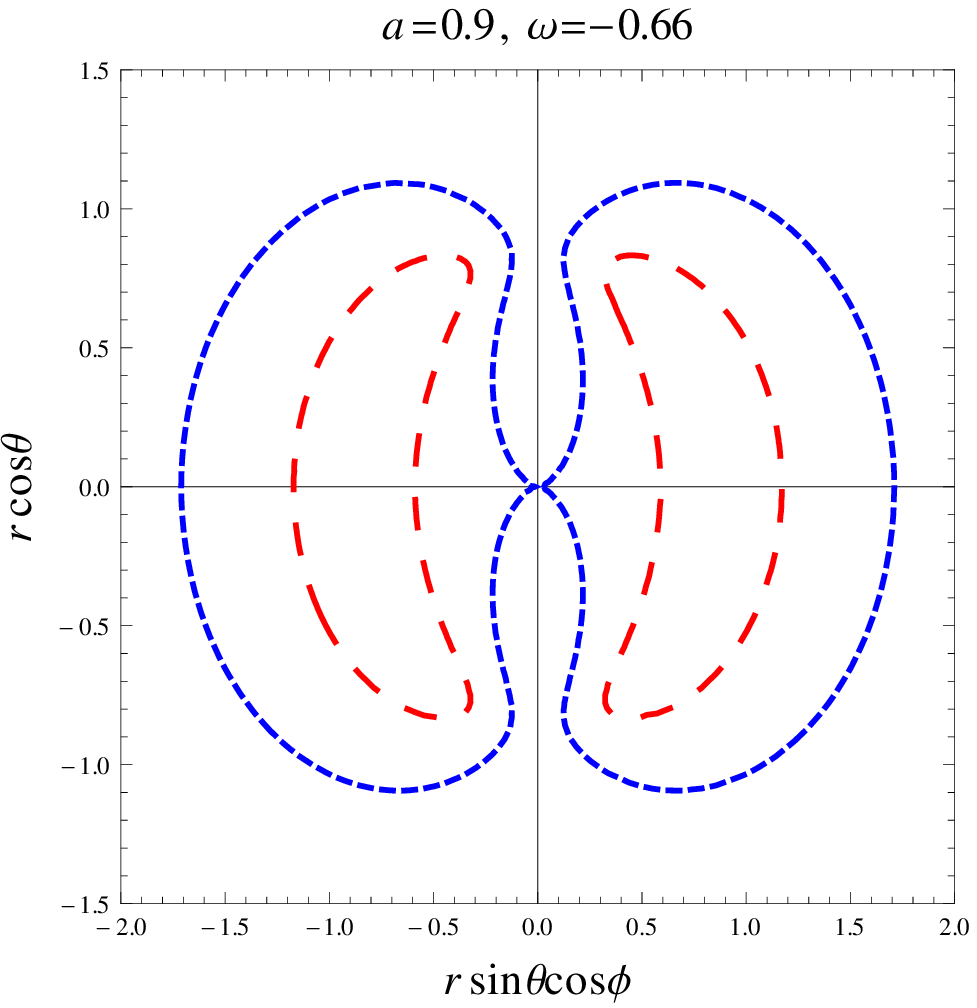}\\
		\includegraphics[scale=0.42]{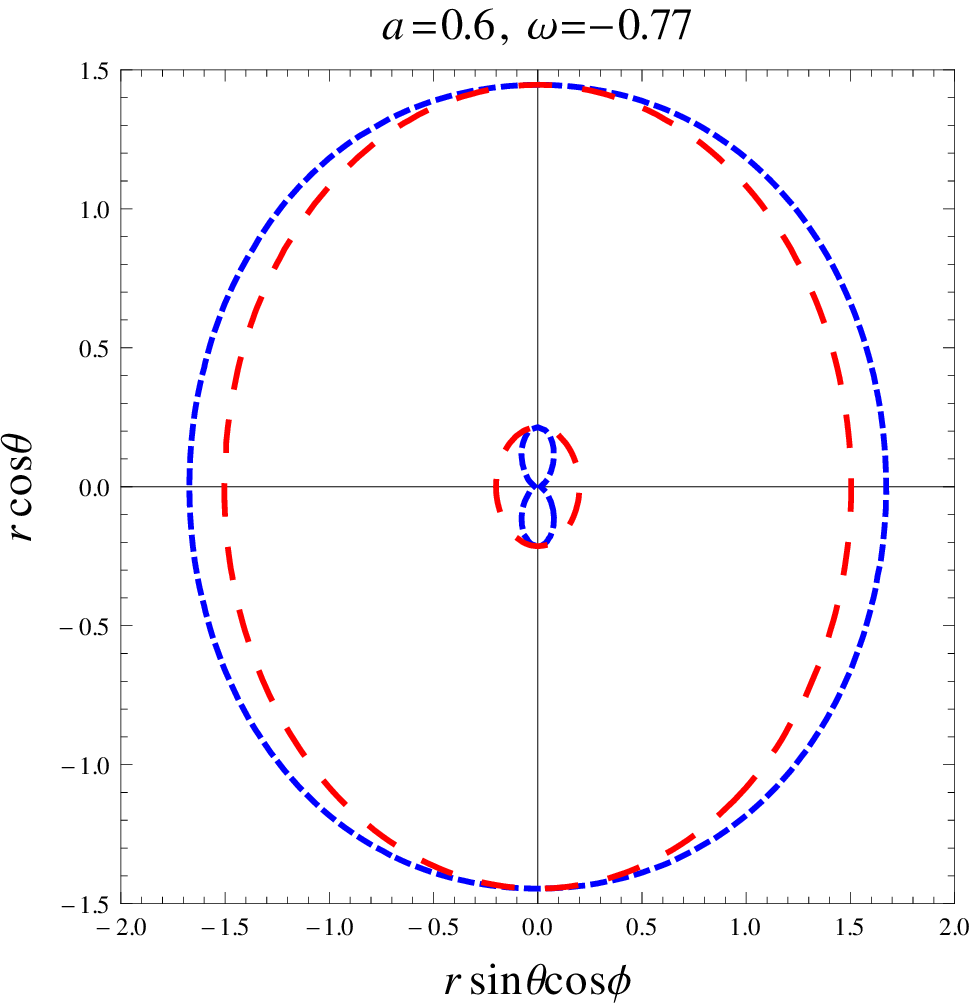}\hspace{-0.2cm}
		\includegraphics[scale=0.42]{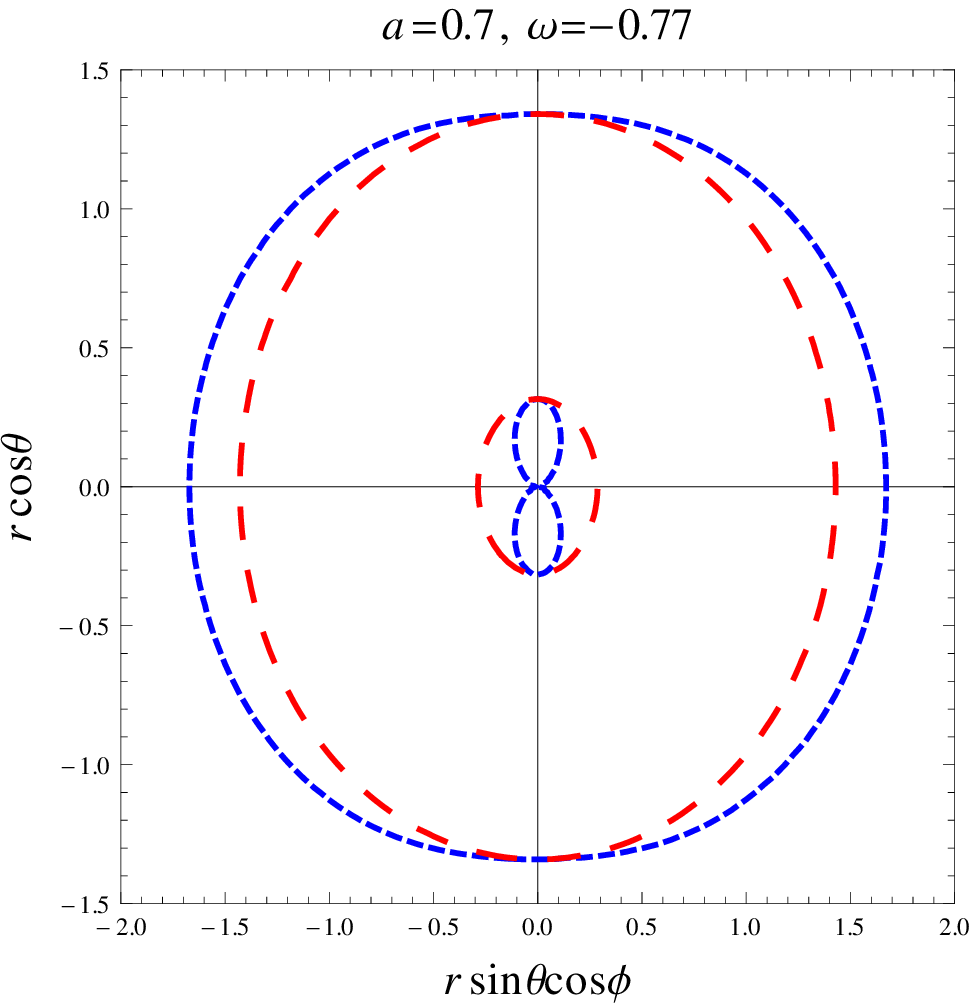}\hspace{-0.2cm}
		\includegraphics[scale=0.42]{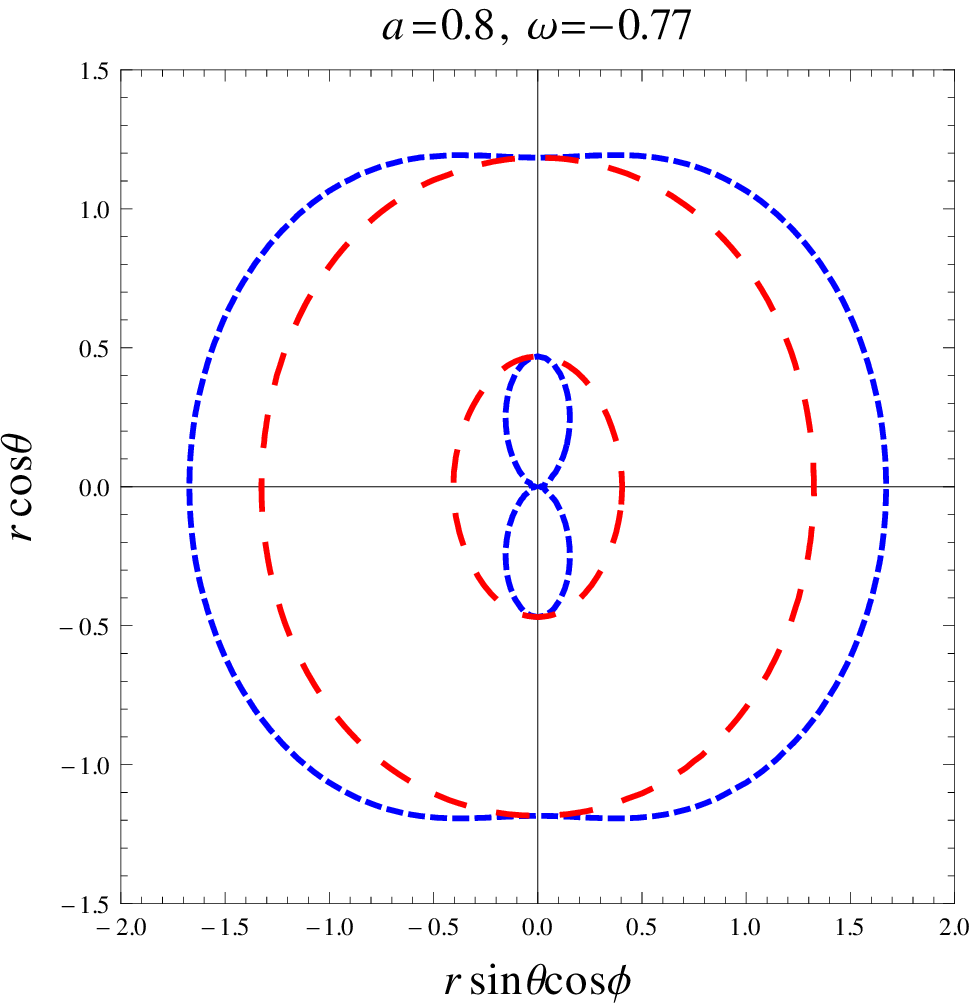}\hspace{-0.2cm}
	   &\includegraphics[scale=0.42]{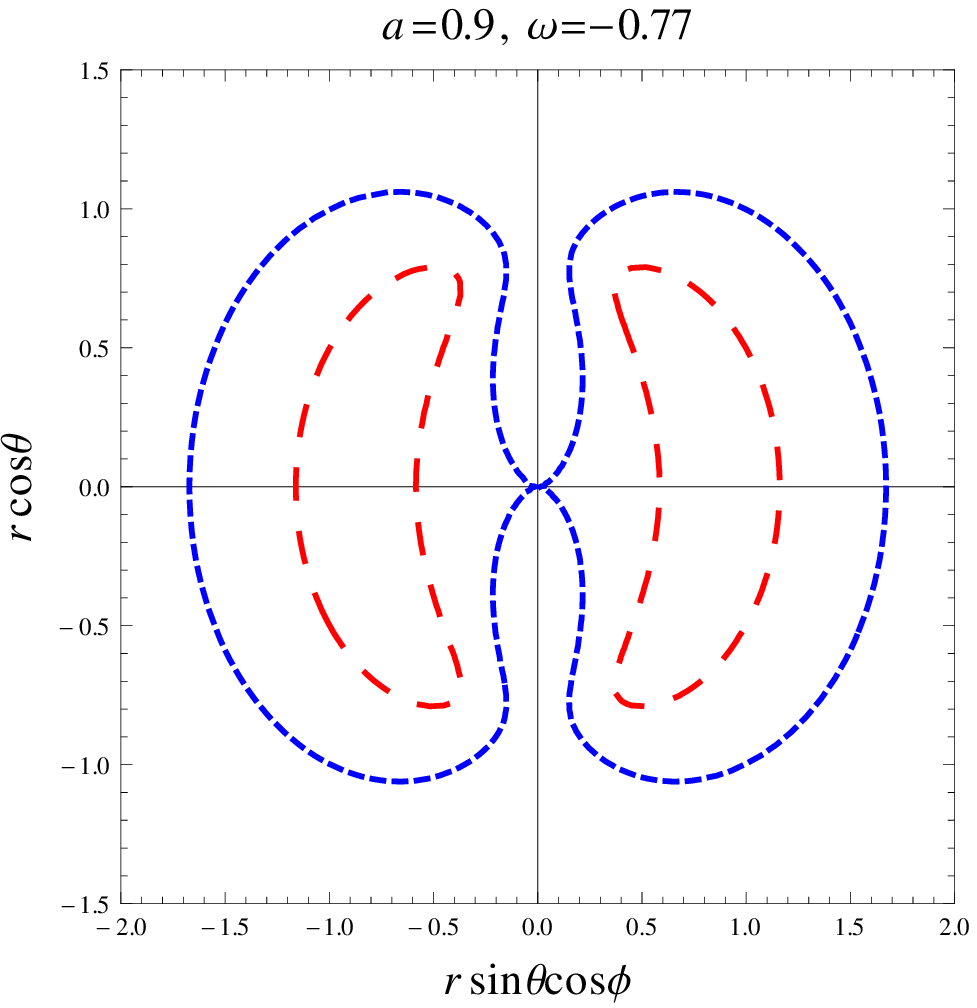}\\
		\includegraphics[scale=0.42]{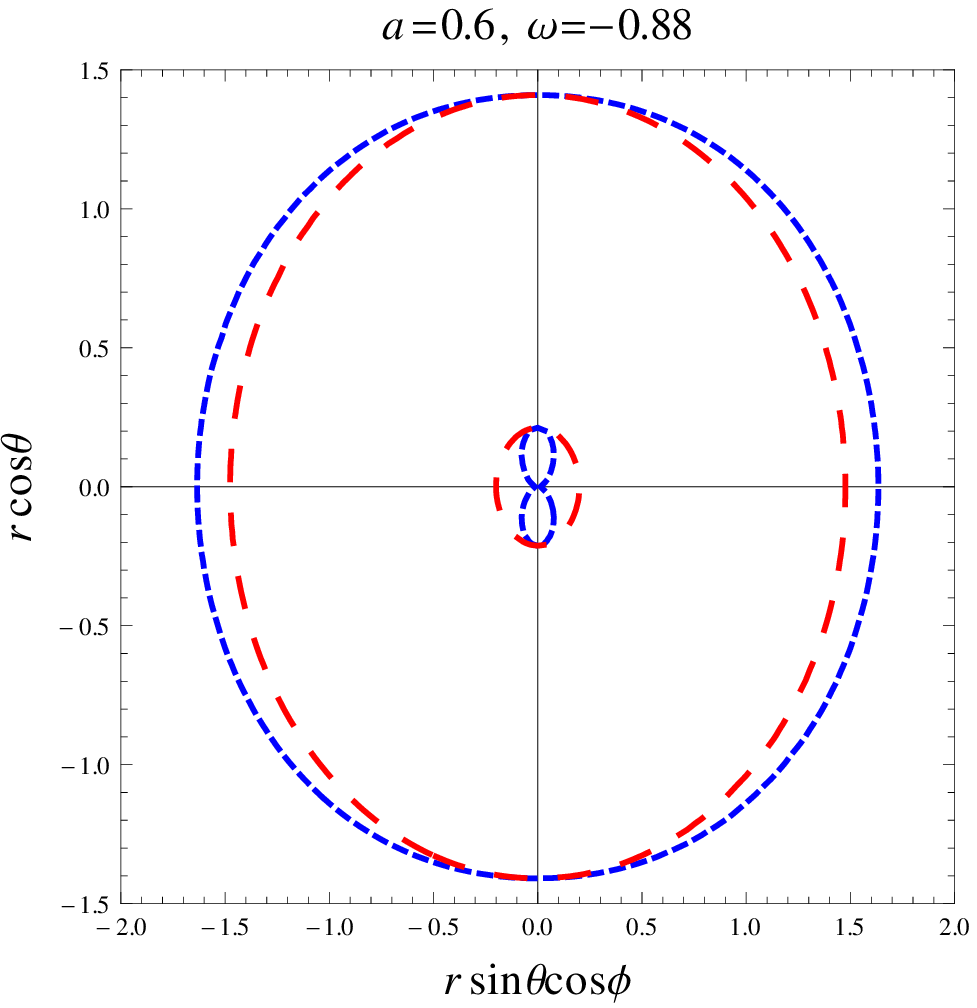}\hspace{-0.2cm}
		\includegraphics[scale=0.42]{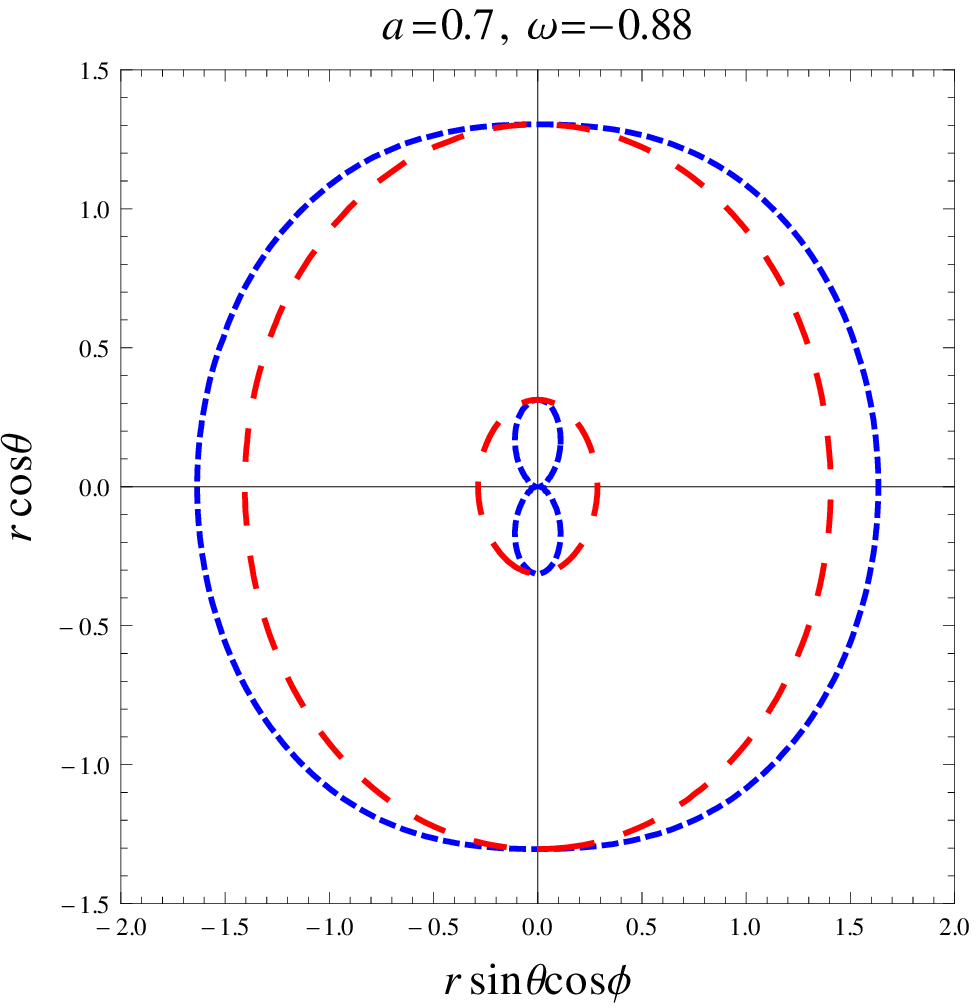}\hspace{-0.2cm}
		\includegraphics[scale=0.42]{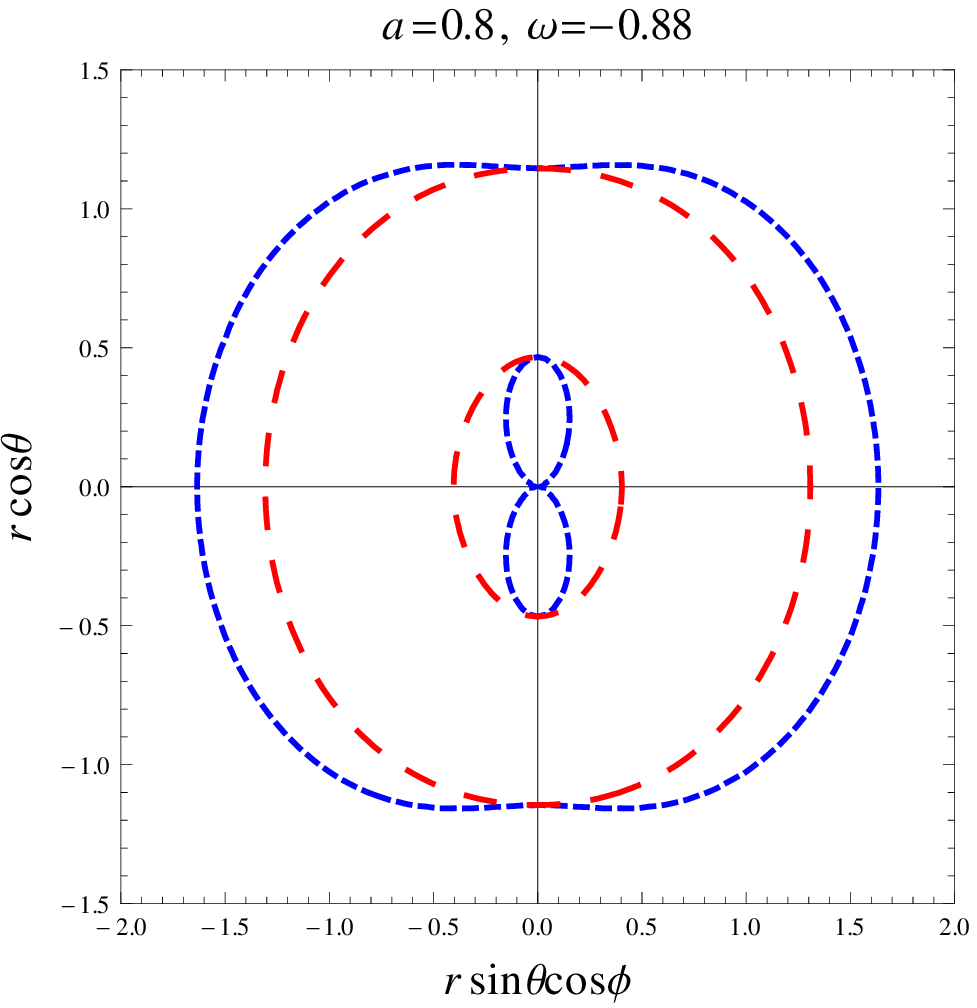}\hspace{-0.2cm}
       &\includegraphics[scale=0.42]{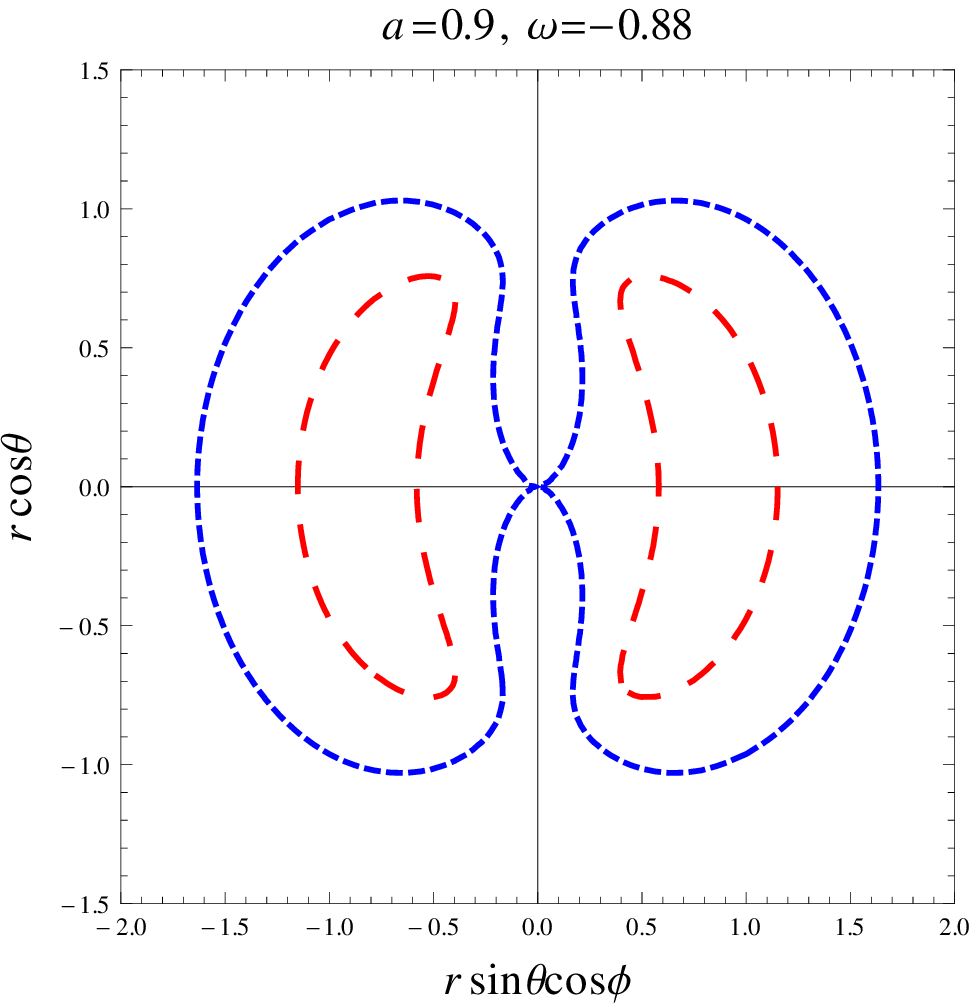}
	\end{tabular}
	\caption{Plot showing the variation of the shape of \textit{ergoregion} in $xz$-plane with parameter $\omega$, for different values of $a$, of the rotating black hole. The blue and the red lines correspond, respectively, to static limit surfaces and horizons. The outer blue line corresponds to the static limit surface, whereas the two red lines correspond to the two horizons}\label{fig3}
\end{figure*}
From the new null tetrad, a new metric is constructed using (\ref{NPmetric}), which takes the form
\begin{eqnarray}
ds^2  &=& \zeta(r,\theta)
du^2+2dudr- 2a\sin^2\theta dr d\phi \nonumber
\\ &&  -\Sigma(r,\theta) d\theta^2
-\left[a^2 \left(\zeta(r,\theta)-2 \right) - \Sigma(r,\theta)\right] \nonumber \\ &&  \times \sin^2\theta
d\phi^2 ~-~2a \left(1-\zeta(r,\theta) \right)\sin\theta^2 du d\phi, \nonumber
 \label{metric2}
\end{eqnarray}
with $\Sigma(r,\theta) = r^2+a^2\cos^2\theta$.

\begin{table}
\caption{The effect of quintessence parameter $\omega$ on the extremal rotation parameter ($a_{E}$) and extremal horizon ($r^{E}$)}\label{table1}
\begin{center}
\begin{tabular}{ | c | c| c | c| c |}
			\hline
           &\multicolumn{2}{c}{$\theta=\pi/4$}  &\multicolumn{2}{c}{$\theta=\pi/2$} \\
            \hline
$\omega$  & $a=a_{E}$              & $r^{E}$  & $a_{E}$              & $r^{E}$  \\
\hline
 -0.50    & 0.9270821091720760   & 0.884066     & 0.9556124089886700   & 0.894288   \\
 -0.66    & 0.9243197545134466   & 0.860520     & 0.9578436527014000   & 0.883037  \\
 -0.77    & 0.9228228129512000   & 0.845995     & 0.9593257246846765   & 0.877225  \\
 -0.88    & 0.9215809590176600   & 0.832531     & 0.9607794407624000   & 0.872548   \\
 \hline
\end{tabular}
\end{center}
\end{table}

Thus we have obtained rotating analogue of the static black hole metric (\ref{solA})
\begin{eqnarray}
ds^2 & = &   \zeta(r,\theta) dt^2 + \frac{\Sigma(r,\theta)}{\Delta(r)} dr^2
\nonumber \\ & &  + 2 \left(1- \zeta(r,\theta) \right) \sin^2 \theta dt d\phi  - \Sigma(r,\theta) d\theta^2
\nonumber \\ & &  - \sin^2 \theta \left[a^2 \left(2 - \zeta(r,\theta) \right)  \sin^2 \theta + \Sigma(r,\theta) \right] d\phi^2.
\end{eqnarray}
In order to simplify the notation we introduce the following quantities
\begin{eqnarray}
\Delta(r) & = & \zeta(r,\theta) \Sigma(r, \theta) +  a^2 \sin^2 \theta
\nonumber \\ & = & r^2 +a^2 - 2 M r - \frac{\alpha}{{\Sigma}^{\frac{3\omega-1}{2}}},
\label{delta2}
\end{eqnarray}
inside the metric and we write down the line element explicitly in Boyer\(-\)Lindquist coordinates
defined by the coordinate transformation $$du = dt - \left(\frac{r^2+a^2}{\Delta}\right) d r,\; d \phi = d \phi' - \frac{a}{\Delta} dr.$$
In above and henceforth, we omit writing the dependent on $\theta$ and $r$ in the function $\Delta$ as well as in $\Sigma$.
Then, this metric could be cast in the more familiar Boyer\(-\)Lindquist coordinates to read as
\begin{eqnarray}
ds^2 && = \frac{\Delta - a^2 \sin^2 \theta}{\Sigma} dt^2 - \frac{ \Sigma}{\Delta }  \, dr^2    \nonumber \\
&&  +  2 a \sin^2 \theta \left(1 - \frac{\Delta - a^2 \sin^2 \theta}{\Sigma} \right) dt \, d \phi
 - \Sigma \, d \theta^2
 \nonumber \\
&&  -  \, \sin ^2 \theta  \left[ \Sigma + a^2 \sin^2 \theta \left(2 - \frac{\Delta - a^2 \sin^2\theta}{\Sigma}\right)   \right]    d \phi^2,
  \label{rotbhq}
\end{eqnarray}
with $ \Delta$ and $\Sigma$ as defined above. This is a rotating black hole metric which for $\alpha=0$ reduces to Kerr black hole, while in the particular case  $a=0$, it reconstruct the Schwarzschild solution surrounded by the quintessence, and for definiteness, we call the metric (\ref{rotbhq}) as rotating quintessence black hole which is stationary and axisymmetric with Killing vectors.

However, like the Kerr metric, the rotating quintessence black hole metric (\ref{rotbhq}) is also singular at $ r = 0 $. The metric (\ref{rotbhq}) generically must have two horizons, viz., the Cauchy horizon and the event horizon. The surface of no return is known as the event horizon. The zeros of  $ \Delta = 0 $ gives the horizons of black hole, i.e., the roots of
\begin{eqnarray}\label{eh2}
 \Delta=r^2 +a^2 - 2 M r -\frac{\alpha}{{\Sigma}^{\frac{3\omega-1}{2}}} = 0.
\end{eqnarray}
This depends on $a,\; \alpha$, $\omega$, and $\theta$, and it is different from the Kerr black hole where it is $\theta$ independent. The numerical analysis of Eq.~(\ref{eh2}) suggests the possibility of two roots for a set of values of parameters which corresponds to the two horizons of a rotating quintessence black hole metric (\ref{rotbhq}). The larger and smaller roots of the Eq.~(\ref{eh2}) correspond, respectively, to the event and Cauchy horizons. An extremal  black hole occurs when $\Delta=0$ has a double root, i.e., when the two horizons coincide. When $\Delta=0$ has no root, i.e., no horizon exists, which mean there is no black hole (cf. Fig.~\ref{fig1} ). We have explicitly shown that, for each $\omega$,  one gets two horizons for $a< a_E$, and when $a=a_{E}$ the two horizons coincide, i.e., we have an extremal black hole with degenerate horizons (Fig.~\ref{fig1} and Table~\ref{table1}). Further, for $a < a_E$, Eq.~(\ref{eh2}) admits two positive roots which are $\omega$ dependent (Fig.~\ref{fig1} and Tables~\ref{table2} and ~\ref{table3}).

\begin{table}
\caption{The Cauchy and event horizons of the black hole for different values of $\omega$ and $a$ ($\theta=\pi/4$)}\label{table2}
  \begin{tabular}{ |c | c c| c c| c c| c c |}
			\hline
&\multicolumn{2}{c}{$\omega=-0.50$}  &\multicolumn{2}{c}{$\omega=-0.66$}  &\multicolumn{2}{c}{$\omega=-0.77$} &\multicolumn{2}{c}{$\omega=-0.88$}  \\
            \hline
$a < a_{E}$  & $r^{-}$ & $r^{+}$ &  $r^{-}$  & $r^{+}$ & $r^{-}$  & $r^{+}$ & $r^{-}$  & $r^{+}$ \\
\hline
0.7          & 0.304099    & 1.46241  & 0.299547  & 1.41561  & 0.297165  & 1.38497  & 0.295200  & 1.35543\\
0.8          & 0.437164    & 1.32997  & 0.430741  & 1.28681  & 0.427068  & 1.25913  & 0.423826  & 1.23282\\
0.9          & 0.671981    & 1.09592  & 0.665427  & 1.05488  & 0.660976  & 1.02987  & 0.656497  & 1.00698\\
\hline
  \end{tabular}
\end{table}

\begin{table}
\caption{The Cauchy and event horizons of the black hole for different values of $\omega$ and $a$ ($\theta=\pi/2$)}\label{table3}
  \begin{tabular}{ |c | c c| c c| c c| c c |}
			\hline
&\multicolumn{2}{c}{$\omega=-0.50$}  &\multicolumn{2}{c}{$\omega=-0.66$}  &\multicolumn{2}{c}{$\omega=-0.77$} &\multicolumn{2}{c}{$\omega=-0.88$}  \\
            \hline
$a < a_{E}$  & $r^{-}$ & $r^{+}$ &  $r^{-}$  & $r^{+}$ & $r^{-}$  & $r^{+}$ & $r^{-}$  & $r^{+}$ \\
\hline
0.7          & 0.289008    & 1.48923  & 0.287523  & 1.45172  & 0.286949  & 1.42732  & 0.286574  & 1.40387\\
0.8          & 0.408981    & 1.37298  & 0.405586  & 1.34310  & 0.404079  & 1.32413  & 0.402983  & 1.30616\\
0.9          & 0.596924    & 1.18918  & 0.588102  & 1.17125  & 0.583599  & 1.16050  & 0.579960  & 1.15067\\
\hline
  \end{tabular}
\end{table}

In the case $\alpha=0$, when the Kerr black hole solution is recovered, there is an event horizon with spherical topology, which is biggest root of the equation $\Delta = 0$, given by
\begin{equation}
r^{\pm} = M \pm \sqrt{M^2 - a^2},
\end{equation}
for $a \leq M$. Beyond this critical value of the spin there is no event horizon and causality violations are present in the whole spacetime, with the appearance of a naked singularity. While the case $\alpha=-e^2 \neq 0$, and $\omega=1/3$ leads to $\Delta = r^2 +a^2 - 2 M r + e^2$ and the roots
\begin{equation}
r^{\pm} = M \pm \sqrt{M^2 - a^2 - e^2},
\end{equation}
correspond to outer and inner horizons of Kerr\(-\)Newman black hole.

In general, as envisaged black hole horizon is spherical and it is given by $\Delta = 0$, which has two positive roots giving the usual outer and inner horizon and no negative roots. The numerical analysis of the algebraic equation $\Delta=0$ reveals that it is possible to find non-vanishing values of parameters $a,\; \omega$ and $\alpha$ for which
$\Delta$ has a minimum, and that $\Delta=0$ admits two positive roots $r^{\pm}$ (cf. Fig.~\ref{fig1}).

The static limit or ergo surface is given by $g_{tt}=0$, i.e.,
\begin{eqnarray}
(r^2+a^2 \cos ^2 \theta) -2 M r - \frac{\alpha}{{\Sigma}^{\frac{3\omega-1}{2}}}  = 0.
\end{eqnarray}
 The behavior of static limit surface is shown in Fig.~\ref{fig2}. The two surfaces, viz. event horizon and static limit surface,  meet at the poles and the region between them give the \textit{ergoregion} admitting negative energy orbits, i.e., the region between $r_+^{EH}\, < r\, < r_+^{SLS}$
 is called \textit{ergoregion}, where the asymptotic time translation
 Killing field $\xi^a=(\frac{\partial}{\partial t})^a$ becomes
 spacelike and an observer follow orbit of $\xi^a$. It turns out that the shape of
 \textit{ergoregion}, therefore, depends on the spin $a$, and parameter
 $\omega$.  Interestingly, the quintessence matter does influence the shape of \textit{ergoregion}  as described in the Fig.~\ref{fig3} when compared with the analogous situation of the Kerr black hole.  Indeed, we have demonstrated that the \textit{ergoregion} is vulnerable to the parameter $\omega$ and \textit{ergoregion} becomes more prolate, and ergoregion area increases as the value of the parameter $\omega$ increases. Further, we find that for a given value of $\omega$, one can find critical parameter $a^C$ such that the horizons are disconnected for $a > a^C$ (cf. Fig.~\ref{fig3}).

 Penrose \cite{rp} surprised everyone when he
  suggested that energy can be extracted from a rotating black hole.
  The Penrose process \cite{rp}  relies on
 the presence of an \textit{\textit{ergoregion}},   which for the solution
 (\ref{rotbhq}) grows with the increase of parameter $\omega$ as well
 with spin $a$ (cf. Fig.~\ref{fig3}). Thus the parameter $\omega$  is
likely to have impact on energy extraction.  It will be also useful to  further study the geometrical properties, causal structures and thermodynamics of this black hole solution.  All these and related issues are being investigated.

In this letter, we have used the complex transformations pointed out by Newman and Janis \cite{nja}, for to obtain rotating solutions from the static  counterparts for the quintessential matter surrounding a black hole.  Interestingly, the limit as $a \rightarrow 0$ is still correct from the point of view of the obtained solution, but it is easy to see that the metric obtained by the complex transformation  is likely to generate additional stress \cite{Ghosh:2014hea, Carmeli,Qw}.   It may be pointed out that in the
general relativity case the source, if it exists, is the same for both
a black hole and its rotating counterpart (obtain by Newman-Janis complex transformations), e.g., the vacuum
for both Schwarzschild and Kerr black holes, and charge for
Reissner–Nordstrom and Kerr-Newman black holes. The
source for the  solution (\ref{solA}) is just quintessence matter, whereas
its rotating counterpart (\ref{rotbhq}), in addition to quintessence matter,
has some additional stresses.

\begin{acknowledgements} We would like to thank SERB-DST
 Research Project Grant NO SB/S2/HEP-008/2014, to  M. Amir for help in
plots.   We also thanks IUCAA for hospitality while this work was being done and ICTP for grant No. OEA-NET-76.
\end{acknowledgements}

\noindent
{\it Note added in proof:} After this work was completed, we learned a similar work by Toshmatov {\it et al} \cite{Toshmatov:2015npp}, which appeared in arXiv a couple of days before.

\end{document}